\documentclass[useAMS,usenatbib]{mn2e}

\usepackage{amsmath}
\usepackage{amssymb}
\usepackage{amsfonts}
\usepackage{amstext}
\usepackage{graphicx}
\usepackage{fancybox}
\usepackage[usenames,dvipsnames]{color}
\usepackage{pstricks}
\usepackage{bm}
\usepackage{soul, color}
\usepackage{latexsym}
\usepackage{pifont}
\usepackage{shorttoc}
\usepackage{subfigure}
\usepackage{textcomp} 
\usepackage{hyperref}
\usepackage{bm}
\usepackage{stfloats}

\newcommand{\lya}{Ly$\alpha$\ }
\newcommand{\hmpc}{\, h^{-1}\, {\rm Mpc}}
\newcommand{\kms}{\, {\rm km}\, {\rm s}^{-1}}

\newcommand{\rpar}{$r_{\parallel}$}
\newcommand{\rperp}{$r_{\perp}$}

\def\sgg#1{ #1} 
\def\jordi#1{ #1} 
\def\changes#1{#1}

\begin{document}

\title[Quasar -- CIV forest cross-correlation with SDSS DR12]{Quasar -- CIV forest cross-correlation with SDSS DR12}
\author[S. Gontcho A Gontcho et al.]
{Satya ~Gontcho A Gontcho$^{1,3}$, Jordi ~Miralda-Escud\'e$^{1,2}$, Andreu ~Font-Ribera$^{3}$,
\newauthor Michael ~Blomqvist$^{4}$, Nicol\'as G. ~Busca$^{5}$, James ~Rich$^{6}$\\ 
$^1$Institut de Ci\`encies del Cosmos, Universitat de Barcelona/IEEC,Barcelona 08028, Catalonia, Spain\\ 
$^2$Instituci\'o Catalana de Recerca i Estudis Avan\c{c}ats, Barcelona,Catalonia, Spain\\
$^3$Department of Physics \& Astronomy, University College London, Gower Street, London, WC1E 6BT, UK\\
$^4$Aix Marseille Universit\'e, CNRS, LAM, Laboratoire d'Astrophysique de Marseille, Marseille, France\\
$^5$LPNHE, CNRS/IN2P3, Universit\'e Pierre et Marie Curie Paris 6, Universit\'e Denis Diderot Paris 7, 4 place Jussieu, \\75252 Paris CEDEX, France\\
$^6$ IRFU, CEA, Universit\'e Paris-Saclay,  F-91191 Gif-sur-Yvette, France}

\date{Accepted ? ; Received ? ; in original ?}

\pagerange{\pageref{firstpage}--\pageref{lastpage}} \pubyear{2018}

\maketitle

\label{firstpage}

\begin{abstract}
 ~\par  We present a new determination of the large-scale clustering of the
CIV forest (i.e., the absorption due to all CIV absorbers) using its
cross-correlation with quasars in the Sloan Digital Sky Survey (SDSS) Data
Release 12 (DR12). We fit a linear bias model to the measured cross-correlation.
We find that the transmission bias of the CIV forest, $b_{Fc}$, at a mean
redshift of $z=2.3$, obeys $(1+\beta_c)b_{F c} = -0.024 \pm 0.003$.
Here, $\beta_{c}$ is the linear redshift space distortion parameter of the CIV
absorption, which can only be poorly determined at $\beta_c=1.1\pm 0.6$ from
our data. The most accurately determined combination marginalized over
$\beta_c$ is $(1+0.44\beta_c)b_{F c} = -0.0170 \pm 0.0014$. The transmission
bias is related to the bias of CIV absorbers and their host halos,
$b_{\tau c}$, through the effective mean optical depth of the CIV forest, which
we estimate at $\bar \tau_c(z) \simeq 0.01$ from previous studies of the CIV
equivalent width distribution. We then find $1< b_{\tau c} < 1.7$, with the
large error arising from uncertainties in $\beta_c$ and $\bar\tau_c$. This CIV
bias is lower than the DLA bias $b_{\rm DLA}\simeq 2$ measured previously
from the cross-correlation of DLAs and the \lya forest, indicating that most
CIV absorbers are hosted by halos of lower mass than DLAs. More accurate
determinations of $\bar \tau_c(z)$ and $\beta_c$ are necessary to check this
conclusion.
\end{abstract}

\begin{keywords}
cosmology -- quasar.
\end{keywords}


\section{Introduction}
\label{sec:Introduction}

~\par The doublet transition of triply ionized carbon (CIV) at
1548.204 \AA\ and 1550.781 \AA\ is one of the strongest and most common
absorption lines probing intergalactic clouds that contain metals
produced in stellar interiors. As a high-ionization species, CIV probes
relatively low density, unshielded gas, and is commonly present in \lya
absorption systems with column densities as low as $N_{\rm HI}\sim
10^{14}\, {\rm cm}^{-2}$ (e.g., \cite{DOdorico2016}), implying that
heavy elements are present in a large fraction of the intergalactic
medium. Studies of metal-absorption systems and their large-scale
distribution, as probed by absorption spectra, can help us understand if
intergalactic metals originated in numerous low-mass halos where the
primordial gas was first able to cool and form stars, or in more massive
galaxies that ejected metal-loaded winds out to large distances and
polluted vast volumes of intergalactic space. CIV systems are easily
observed in the epoch when the global star formation rate in galaxies
peaked, at a redshift $z \sim$ 2 to 3, (e.g.,
\sgg{\cite{Schaye2003,Hopkins2006,Rauch1997,Boksenberg2015}}). Therefore, they can
crucially contribute to our understanding of galaxy formation and
evolution by probing gas in the process of cooling and accreting onto
galaxies, as well as gas flowing out in winds.

~\par Studying the large-scale clustering strength of CIV systems, in
particular measuring their cross-correlation with other tracers of
known auto-correlation, can help reveal the origin of these systems.
On large scales, the clustering of any tracer can be described by linear
theory, which depends on a linear bias factor that is monotonically
related to the halo mass they reside in (e.g., \cite{Cole1989,Tinker2010}).
Depending on the model assumed for the metal enrichment process, we
expect different values for the clustering strength: if most CIV systems
originated in stars forming in very low-mass halos at all epochs, the
gas should reflect the low bias factor of the smallest galaxies, whereas
if the CIV gas was expelled to the IGM in winds from massive
star-forming galaxies, then their bias factor would be high, reflecting
the highly biased nature of the most massive dark matter halos. Detailed
hydrodynamic simulations of galaxy formation can model the absorption
line systems arising during the process of accretion of gas into
galactic halos and ejection by winds, making crucial predictions for
testing our understanding of how galaxies form (see \cite{Bird2016} for
a recent study).

~\par Previous studies have estimated the clustering strength of
individual CIV absorbers between redshift 1.5 and 4.5 (e.g,
\cite{Vikas2013,Lundgren2013}). In this paper, instead of using
individually identified CIV absorption systems, we use spectral flux
fluctuations blueward of the quasar CIV emission line as a continuous
absorption field probing carbon enriched gas in the Universe, similarly
to the studies of the \lya forest for atomic hydrogen. We refer to this
continuous absorption field as the \textit{CIV forest}. This approach
has the advantage of including all the absorption systems, whether or
not they are individually detected, and avoiding any dependence on the
detection method. Using flux fluctuations directly has been done so far
only for the \lya forest, for which there was also a long debate about the
nature of the absorption line systems as individual clouds or part of
the intergalactic medium (e.g.,
\cite{Bahcall1969,Lynds1971,Sargent1980,Miralda1996,Rauch1998}). 
The approach of measuring the \lya forest clustering in terms of
the power spectrum of the continuous absorption field, starting with
\cite{Croft1998,Croft1999,Mcdonald2000,Croft2002,Mcdonald2006},
has generally been very successful.
In the linear regime, the over-density of a tracer $\delta_t$ in real
space relates to the over-density in mass $\delta$ through the 
bias factor $b$ as $\delta_t = b\cdot\delta $; it follows that 
the observed \lya forest correlation function should be equal to the
mass autocorrelation times the square of the mean bias factor of the
\lya transmission fluctuation, with the appropriate modifications of
the well known linear redshift space distortions \citep{Kaiser1987}.
Measurements of this \lya autocorrelation have allowed the measurement of the value 
of the \lya bias and redshift space distortion parameters
\citep{Slosar2011,Slosar2013,Busca2013,Delubac2015,Blomqvist2015,Bautista2017}. 
In addition to these measurements, the quasar bias has also been 
measured from its auto-correlation \citep{Croom2004,Myers2006a,Myers2006b,
Coil2007,Shen2006,Ross2009,Shen2008,White2012,Eftekharzadeh2015,Laurent2017}
and from its cross-correlation with the \lya forest
\citep{Font2013,Font2014,Helion2017}, using the fact that the
cross-correlation between two different probes is proportional to the
product of bias factors of the two probes.

~\par Similarly, here we measure the cross-correlation of quasars and the
CIV forest at a mean redshift of $z=2.3$ using the final data release of
the Baryon Oscillation Spectroscopic Survey (BOSS) from the Sloan Digital
Sky Survey 3rd edition (SDSS-III), inferring a CIV absorption
fluctuation bias. Although the CIV forest is optically much thinner than the \lya
forest, and is therefore better described as individual absorbers for
many applications, the alternative of directly measuring continuous
fluctuations is still worth investigating.

~\par \jordi{
The bias of the CIV forest large-scale transmission
fluctuations, in the same way as for the Ly$\alpha$ forest, depends not
only on the bias of the absorbing clouds or
their host halos, but also on the effective optical depth,
$\bar \tau_c(z)$,
quantifying the mean CIV transmission as a function of redshift,
$\overline F_c(z)=\exp(-\bar\tau_c(z))$.} We will see that the directly
measurable fluctuation amplitude of CIV transmission fluctuations is
proportional to $\bar\tau_c$, meaning that we need independent
information on $\bar\tau_c(z)$ in order to translate 
the measured CIV forest bias to a bias of the absorbers, equal to that
of their host halos. 

~\par Measuring the bias factor of CIV forest transmission fluctuations is
also useful to estimate the level of CIV contamination in clustering
measurements of the \lya forest, one of the main systematics studied
in the recent \lya analyses from the BOSS collaboration
\citep{Bautista2017,Helion2017}. An additional motivation to study the
clustering of the CIV forest is to use it as a new tracer of the
underlying mass density field that can probe the initial power spectrum,
for example to measure the scale of Baryon Acoustic Oscillations
\cite{Pieri2014}.

~\par In parallel to the present work, \cite{Blomqvist2018} have completed 
a similar investigation including the larger data set of eBOSS.
\jordi{We compare and discuss the implications of the results
in section \ref{ssec:comp}.}

~\par We start introducing the data samples used for this study in
section \ref{sec:data}. 
In section \ref{sec:getDCIV} we describe our procedure
to obtain the transmission fluctuation in the CIV forest. Our method
to infer the cross-correlation and fit a model is explained in
section \ref{sec:cross}.
In section \ref{sec:results} we present our measurement of the
quasar-CIV forest cross-correlation and the inferred CIV bias factor. 
We discuss the interpretation of our results in terms of the
bias factor of the CIV absorber host halos in section \ref{sec:ccl},
and we conclude in section \ref{sec:conc}.
Throughout this work, we use a flat $\Lambda$CDM cosmology with $h=0.678$, 
$\Omega_m = 0.268$, $\Omega_b$=0.049,  $\sigma_8 = 0.815$ and $n_s = 0.968$, 
the best fitting model to the CMB anisotropy power spectrum
found in \cite{Planck2015}.


\section{Data sets}\label{sec:data}

~\par This paper uses the public twelfth Data Release (DR12, \cite{DR122015}) of the SDSS-III 
Collaboration (\cite{Eisenstein2011,Bolton2012,Gunn1998,Fukugita1996,Gunn2006,York2000}), 
encompassing the entire 5 years of observations of the Baryon Oscillation Spectroscopic Survey (BOSS, \cite{Dawson2013}).

\subsection{Quasar catalog}\label{ssec:Qselec}

~\par
\begin{figure*}
\includegraphics[width=\textwidth]{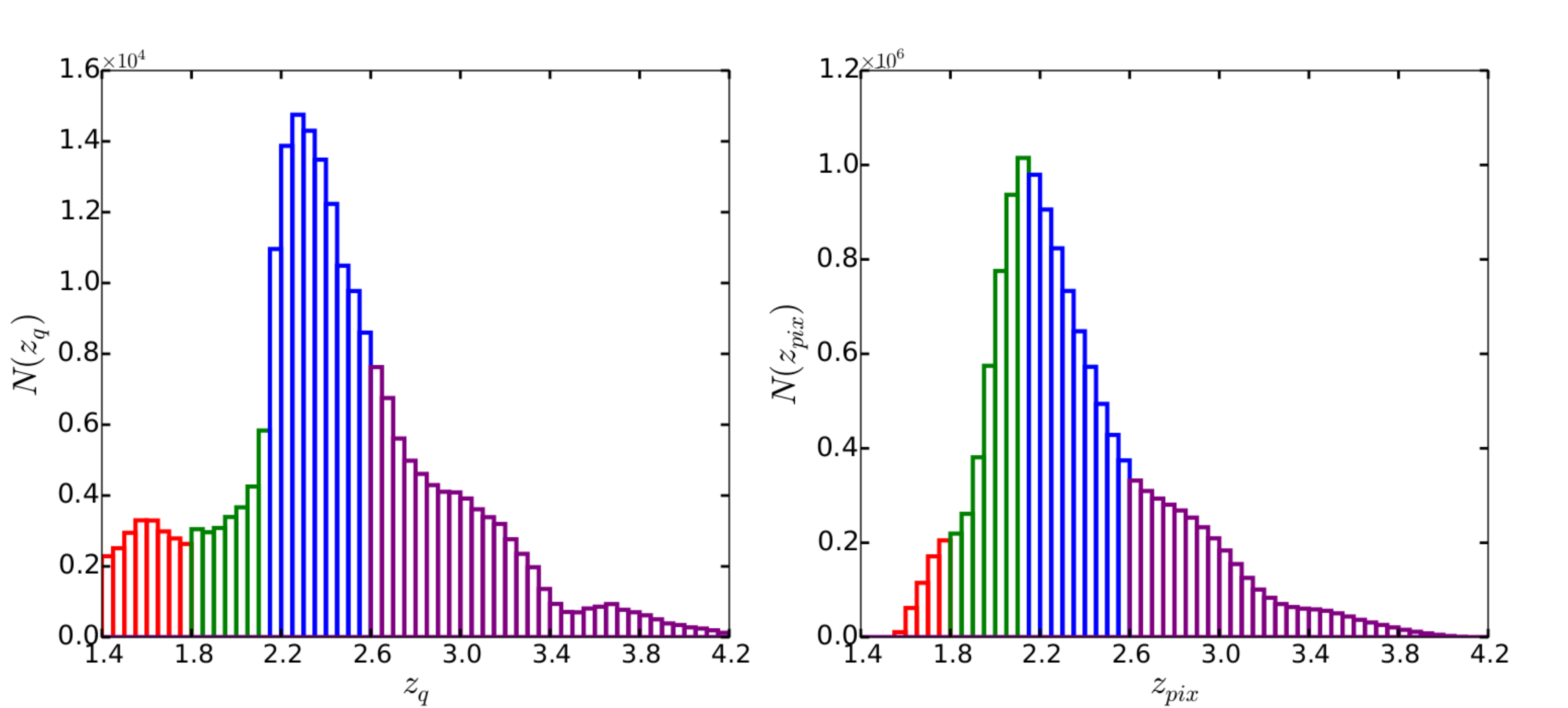}
\caption{Left panel presents the distribution of quasar redshift, and
the right panel presents the distribution of forest pixel redshifts
(calculated with a mean CIV absorption wavelength at $1549.06$ \AA).
Colors indicate the redshift intervals (described in table \ref{tab:z})
used throughout our analysis: $1.4 \leq z < 1.8$ (red),
$1.8 \leq z < 2.15$ (green), $2.15 \leq z < 2.6$ (blue), and
$2.6 \leq z \leq 4.2$ (purple).
}
\label{fig:hist}
\end{figure*}

 The DR12 quasar catalog\footnote{\href{http://www.sdss.org/dr12/algorithms/boss-dr12-quasar-catalog/}{http://www.sdss.org/dr12/algorithms/boss-dr12-quasar-catalog/}},   
described in \cite{Paris2016}, contains 297,301 quasars which were
targeted for spectroscopy using the target selection procedure presented
in \cite{Ross2012}, which combines several algorithms to identify
candidates described in \cite{Richards2009,Kirkpatrick2011,Yeche2010,Bovy2011}.
We impose the redshift cut $1.4 \leq z_q \leq 4.2$, which reduces the catalog to 231,312 quasars.
We use the quasar redshifts obtained with the Principal Component Analysis method, as
described in \cite{Paris2012,Paris2014,Paris2016}.
The redshift distribution of the quasar catalog is illustrated in Figure
\ref{fig:hist} (left panel).

\subsection{CIV absorption spectra sample}\label{ssec:CIVselec}

~\par We now define the set of spectra that we use for measuring the CIV
absorption, which are different from the set of quasars described above.
Starting again from the 297,301 quasars in the catalog of
\cite{Paris2016}, we first apply a cut to eliminate quasars with
detected broad absorption lines (BALs).
BALs appear in quasar spectra when jets pointing close to our direction
launch matter at velocities of thousands of $\kms$, producing absorption
features blueward of the \lya and other emission lines that can be
confused with intergalactic clouds between the quasar and us. The
broadness of these lines, caused by the large velocity dispersion of the
outflowing gas, is used to identify BALs and classify quasars according
to the "balnicity index" (BI), introduced in \cite{weymann1991}.
We eliminate all quasars that were flagged as BALs in the visual
inspection. This leaves 265,889 quasars with a BI equal to zero in
our sample. 
  
~\par We define the CIV forest to be the interval $1420$ \AA $\, \leq \lambda_{RF} <
1520$ \AA, where $\lambda_{RF}$ is the quasar rest-frame wavelength.
\sgg{This avoids the region close to the quasar CIV
emission line, which has a variable profile and is more sensitive to
the presence of undetected BALs, and also the region that is affected by
the SiIV quasar emission line and SiIV forest lines at lower
wavelengths (see figure \ref{fig:check})}. We use the average wavelength of the
unsaturated CIV doublet, $\lambda_c = 1549.06$ \AA, to convert pixel
wavelengths to redshifts in the CIV forest. 
  
~\par We apply a redshift cut for the quasar redshift of the spectra
used to measure the CIV absorption, set to $1.8 \leq z_{q}  \leq 4.2$, leaving 182,566 quasars. 
The lower limit is set by the requirement to have a margin on the blue side of the CIV forest 
to estimate the quasar continuum with the method that is
described below in section \ref{ssec:contfit}, and we do not use pixels below 
3600 \AA\ as the signal-to-noise ratio gets degraded for bluer wavelengths. 
As a result, only complete CIV forests are included in our work sample.
The upper limit is set
to the same maximum quasar redshift of 4.2 used above, beyond which the
BOSS surface density of quasars is not sufficient to be useful for this
study.

~\par We remove pixels in which the variance of the co-added
sky-subtracted sky fibers is significantly higher than 
in neighboring pixels. The DR12 sky mask provides a list of these
observed-frame wavelengths\footnote{\href{https://github.com/igmhub/picca/blob/master/etc/dr12-sky-mask.txt}{https://github.com/igmhub/picca/blob/master/etc/dr12-sky-mask.txt}
}. To apply the mask, we
remove any pixels with an observed wavelength, $\lambda$, in the range 
\begin{equation}\label{eq:sm12}
 {\rm abs} \left[ 10^4\cdot\log_{10}(\lambda / \lambda_{mask}) \right]
 \leq  m  ~,
\end{equation}
for any $\lambda_{mask}$ in the list, where $m$ is the margin. We use a
margin of $1.5$.
Note that pixels in the BOSS co-added spectra have a wavelength width of
$\Delta\log_{10}\lambda = 10^{-4}$.

~\par The final cut we make on the spectra used for measuring the CIV
absorption is related to our method to determine a quasar continuum.
We define for this purpose two spectral zones surrounding the CIV
forest that avoid the SiIII, SiIV and CIII emission lines: the first
zone, referred to as $\Lambda_1$, is $1280$ \AA $ \leq \lambda_{RF} \leq 1380$ \AA,
and the second zone $\Lambda_2$ is $1575$ \AA $ \leq \lambda_{RF} \leq 1860$ \AA,
where $\lambda_{RF}=\lambda/(1+z_q)$ is the quasar rest-frame wavelength. 
A spectrum is retained in our sample if the following condition for the
signal-to-noise is satisfied in these two regions: 
\begin{equation}
{1\over n_{pc} } \sum_i\left( {f_{si}/\sigma_{si}} \right)\, \geq 2 ~,
\end{equation}
where the sum is done over all pixels in regions $\Lambda_1$ and
$\Lambda_2$, $f_{si}$ and $\sigma_{si}$ are the observed flux and noise in
a pixel $i$ of spectrum $s$, and $n_{pc}$ is the total number of pixels in 
the sum, in both regions $\Lambda_1$ and $\Lambda_2$. After the spectra that do
not reach this minimum average signal-to-noise are eliminated, our
sample is reduced to 140,813 lines of sight, with the redshift
distribution of pixels in the CIV forest region
($1420$ \AA $ \leq \lambda_{RF} < 1520$ \AA) shown in Figure \ref{fig:hist}
(right panel).


\section{From flux in the BOSS spectra to the CIV transmission
 fluctuation $\delta_{c}$}\label{sec:getDCIV}

\subsection{Continuum fitting}\label{ssec:contfit}

~\par To obtain the fraction of absorbed flux due to the presence of CIV
systems in each spectrum we need a \emph{continuum} model for each
quasar, which is the quasar flux that would be observed in the absence
of any intervening absorption. We use a weighted average of the sample
of CIV absorption spectra defined in section \ref{ssec:CIVselec} for our
continuum model, with weights that are proportional to the square of the
signal-to-noise. We first derive normalizing coefficients of the flux
for each spectrum $s$ in our sample, in our two normalizing intervals
\begin{equation} \label{eq:c1}
c_{s1} = \sum_{i\in \Lambda_1} \left( w_{si} f_{si} \right) \Big/ 
\sum_{i\in \Lambda_1} \left( w_{si} \right) ~,
\end{equation}
\begin{equation} \label{eq:c2}
c_{s2} = \sum_{i\in \Lambda_2} \left( w_{si} f_{si} \right) \Big/ 
\sum_{i\in \Lambda_2} \left( w_{si} \right) ~,
\end{equation}
where the sum is over all pixels $i$ of a spectrum $s$ that belong to
the $\Lambda_1$ region for $c_{s1}$, and the $\Lambda_2$ region for
$c_{s2}$. The observed flux at every pixel is $f_{si}$ and the weight is
set to the inverse variance, $w_{si}=1/\sigma_{si}^2$ (where 
$\sigma_{si}$ is the noise in every pixel as estimtated by the SDSS pipeline). 
The linear regression to the flux in each spectrum from these mean values is,
\begin{equation}
 L_{si}= c_{s1} + { c_{s2} - c_{s1} \over \lambda_2 - \lambda_1 } \,
  (\lambda_i - \lambda_1) ~,
\end{equation}
where $\lambda_i$ is the observed wavelength of pixel $i$, and
$\lambda_1$ and $\lambda_2$ are the mean observed wavelengths of
regions $\Lambda_1$ and $\Lambda_2$ (with the pixels weighted the same
way as $c_{s1}$ and $c_{s2}$ in equations \ref{eq:c1} and  \ref{eq:c2}).
The normalized flux in each
spectrum, after dividing by the mean value and removing the spectral
tilt between regions $\Lambda_1$ and $\Lambda_2$, is
\begin{equation} \label{eq:ren}
\hat{f}_{si} = \frac{f_{si}}{L_{si}} ~,
\end{equation}
and the weight assigned to this normalized flux is set proportional to
the square of the signal-to-noise,
\begin{equation}\label{eq:w1}
\hat{w}_{si}=L_{si}^2 w_{si} ~.
\end{equation}

~\par
\begin{figure*}
\includegraphics[width=\linewidth]{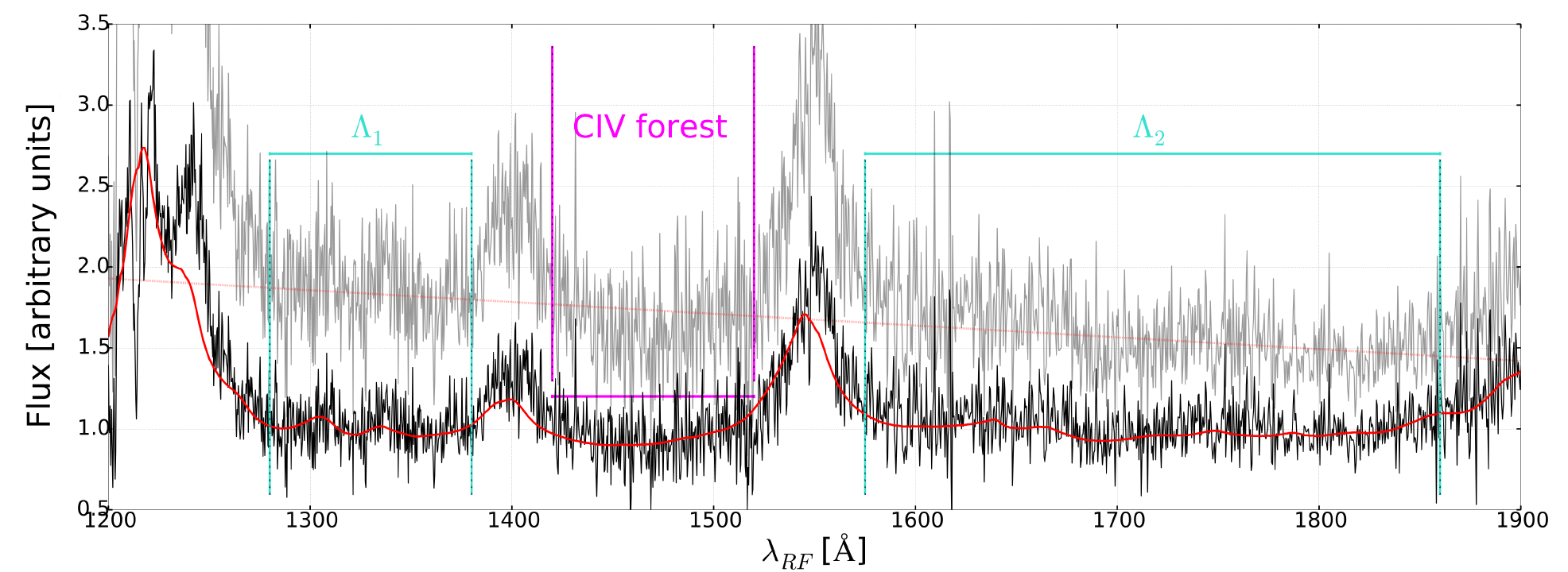}
\caption{Illustration of our procedure to estimate the mean quasar
continuum in the CIV forest region, for the BOSS quasar spectrum at
$RA=1.490^{\circ}$, $DEC=13.582^{\circ}$, $z=2.896$,
reference PLATE = 6177, MJD = 56268, FIBER = 55 and THINGID=232251732. 
The gray spectrum is the observed spectrum $f_{s}$. The red dashed line
is the linear regression to the mean flux in regions $\Lambda_1$ and
$\Lambda_2$ (shown by vertical dotted turquoise lines), and the black
spectrum $\hat{f}_{si}$ is the normalized flux after dividing by the
linear regression. The red solid line is the mean quasar continuum
introduced in equation \ref{eq:cont}. The magenta vertical lines mark the
CIV forest region we use.}
\label{fig:check}
\end{figure*}

 Finally, we compute the weighted mean quasar continuum at a given 
restframe wavelength, $C_i$, as 
\begin{equation}\label{eq:cont}
C_i=\sum_{s} \left(\hat{w}_{si}\cdot \hat{f}_{si} \right) \Big/
 \sum_{s} \hat{w}_{si} \,,
\end{equation}
where the sum is now over all quasar spectra.
Figure \ref{fig:check} illustrates this normalizing procedure for an
example spectrum.

~\par \jordi{Note that by using normalization regions outside of the CIV 
forest, as highlighted on figure \ref{fig:check}, 
we do not systematically remove any fluctuations and we avoid having
to correct for continuum distortions as in other
BOSS analyses (e.g., discussion on distortion matrix in 
\cite{Helion2017} and \cite{Bautista2017}).}

\subsection{Flux transmission fraction in the CIV Forest}\label{ssec:det}

~\par The flux transmission fraction can now simply be
defined as $\hat{f}_{si}/C_i$ in every pixel. However, the optimal
weights to be used for every spectrum to measure the CIV
cross-correlations may vary from those in equation \ref{eq:w1},
depending on how strong the intrinsic sampling variance is compared to
the observational noise, and this can change the mean value of the
transmission fraction. We therefore define the flux transmission
fluctuation including a global correcting factor $m_i$
(similar to $\bar{F}(z)$ used for analysis of \lya forest correlations in
\cite{Bautista2017}),
\begin{equation}\label{eq:dqciv}
\delta_{c,si}=\frac{\hat{f}_{si} }{C_i\, m_i} - 1 ~.
\end{equation}
Note that while $C_i$ is a function of restframe wavelength, $m_i$ is a 
function of observed wavelength.

~\par The correction $m_i$ is to ensure that the mean fluctuation is equal to
zero when averaged over all the spectra in any given sample, and depends
on new weights $W_{si}$ that we use for measuring correlations of
$\delta_{c,si}$:
\begin{equation}\label{eq:dciv2}
m_i = \sum_{s} \left (W_{si} \frac{\hat{f}_{si} }{ C_i} \right) \Big/ 
\sum_{s} \left( W_{si} \right) ~.
\end{equation}
In this paper, we shall use weights taking into account an intrinsic
sampling noise at every pixel of $\sigma_{int}$, arising from the
shot noise of the CIV absorbers themselves, plus any
other metal-lines or intrinsic variations in the quasar spectra. This
sampling variance implies that spectra at very high signal-to-noise
should not be weighted in proportion to the inverse variance arising
from the observational error, because the intrinsic variance dominates
the contributed error to any correlation measurement. The optimal weight
to use is
\begin{equation}\label{eq:w2}
W_{si}=\left[\left(C_i^2\cdot \hat{w}_{si}\right)^{-1} + \sigma_{int}^2\right]^{-1}\,.
\end{equation} 
The correction factor $m_i$ ensures that
\begin{equation}\label{eq:flatn}
\sum_{s} W_{si} \delta_{c,si} = 0 \, ,
\end{equation}
which guarantees that measured correlations of $\delta_{c,si}$ go to
zero in the limit of large separations.

~\par In the limit when the CIV forest absorption is mainly caused by
saturated systems that absorb most of the flux in one pixel,
the optimal intrinsic dispersion is $\sigma_{int} \sim
(1-\overline F_c)^{1/2}$,
where $\overline F_c$ is the mean transmission fraction. In practice the
optimal value of $\sigma_{int}$ is a little smaller because lines are
not fully saturated. We shall estimate
below from the observations of \cite{Cooksey2013} a value
$1 - \bar{F}_{c}\simeq 0.01$, and we will generally use a characteristic
intrinsic dispersion $\sigma_{int} = 0.08$, which we have found to be
near optimal by trying several values of this parameter. 


\section{Quasars-CIV Forest cross-correlation }\label{sec:cross}

 This section describes how we measure the cross-correlation of the
CIV transmission fluctuation $\delta_{c}$ with quasars, and how we fit
it to a linear analytic model.

\subsection{Sub-samples}\label{ssec:subsamples}

\begin{figure*}
\includegraphics[width=\textwidth]{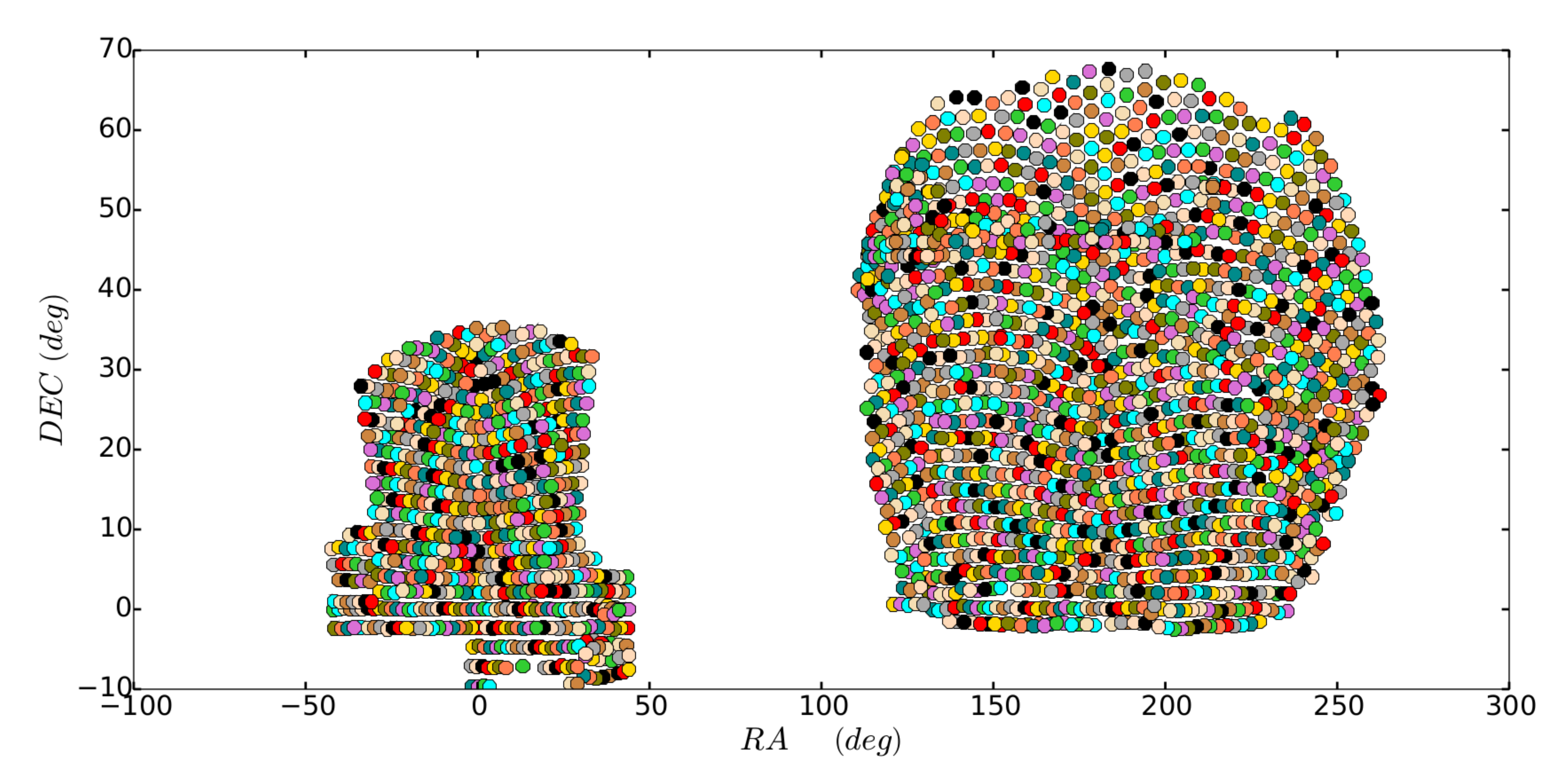}
\caption{DR12 footprint, in J2000 equatorial coordinates, of the 2,370
plates, equal to our sub-samples, shown in different colors as a
visualization aid. The left (right) area is the South (North) Galactic
Cap. The fixed dot size representing each plate is arbitrary and
indicates only the center of the plate, not its actual size.}
\label{fig:plates}
\end{figure*} 

~\par The DR12 sample of CIV spectra described in \ref{ssec:CIVselec} is
divided into 2,370 sub-samples, corresponding to the 2,370 observational
plates of the final DR12 data release of SDSS--III BOSS. Using directly
the observational plates has the advantage that the sub-samples are of
similar sizes and shapes. The distribution of these plates in the sky in
equatorial coordinates is shown in Figure \ref{fig:plates}. These
sub-samples will be used to estimate the cross-correlation and its
covariance matrix via the bootstrap method, combined with a smoothing
mechanism to reduce sampling noise.

~\par Apart from the sky sub-sampling, the CIV spectra are also divided into
four redshift intervals in which we separately measure the
cross-correlation. Table \ref{tab:z} specifies the redshift range,
the redshift average and the proportion of pixels in the CIV forest in
each of the four intervals.

\begin{table}
\begin{center}
\begin{tabular}{cccc}
\hline
Name & $z-$range & $\langle z \rangle$ & Proportion \\
\hline
$\mathbb{F}$ & 1.4 - 4.2 & 2.2981 & 100 \% \\
v$\mathbb{L}$ & 1.4 - 1.8 & 1.7225 & 4 \% \\
$\mathbb{L}$ & 1.8 - 2.15 & 2.0325 & 20 \% \\
$\mathbb{M}$ & 2.15 - 2.6 & 2.327 & 63\% \\
$\mathbb{H}$ & 2.6 - 4.2 & 2.8519 & 13\% \\
\hline
\end{tabular}
\caption{Redshift intervals used in this analysis, with their mean
redshift and the proportion of the whole sample of CIV spectra included
in each one. The redshift interval names are: $\mathbb{F}$ for
\emph{full}, v$\mathbb{L}$ for \emph{very low}, $\mathbb{L}$ for
\emph{low}, $\mathbb{M}$ for \emph{medium} and $\mathbb{H}$ for
\emph{high}.} 
\label{tab:z}
\end{center}
\end{table} 

\subsection{Estimator for the CIV transmission - quasar cross-correlation}
\label{ssec:estimator}

~\par Our method to estimate cross-correlations closely follows that of
\cite{Font2012}, where the cross-correlation of DLAs with the \lya forest 
was measured. 
The comoving separation between any CIV spectral pixel and a quasar is 
expressed in terms of its perpendicular and parallel components,
$r_\perp = D_A(z)(1+z)\, \Delta \theta $ and
$r_\parallel = c/H(z)\, \Delta z$, which we calculate from the angular and
redshift separations $\Delta\theta$ and $\Delta z$, and the angular
diameter distance $D_A(z)$ and Hubble constant $H(z)$ at redshift $z$.
We measure the cross-correlation in bins of $4 \hmpc$, both in the
parallel and perpendicular separation, out to maximum values of
$64\hmpc$. There are therefore a total of $16\times 32 = 512$ bins,
since the separation $r_\perp$ is positive and has 16 bins, and \rpar\
may have positive or negative sign and has 32 bins. The sign of \rpar\
is defined to be positive when the CIV pixel is at higher redshift than
the quasar.

~\par We designate as $\xi_{\mathcal{P},A}$ the measured cross-correlation of
quasars and the CIV transmission fluctuation $\delta_c$ in a plate (or
sub-sample) $\mathcal{P}$, and a bin $A$ in (\rpar , \rperp). The
set of pairs of a quasar and a spectral pixel $i$ contributing to
$\xi_{\mathcal{P},A}$ satisfy two conditions: (a) the pixel $i$ belongs
to a spectrum $s$ located in plate $\mathcal{P}$, and (b) the separation
between the quasar and pixel $i$ is within the bin $A$. The
cross-correlation is computed as
\begin{align}\label{eq:xi_sub}
& \xi_{\mathcal{P},A} = \sum_{s\in\mathcal{P}, i\in A}
  \left(  W_{si}\, \delta_{c,si} \right)  \Big/ \, W_{\mathcal{P},A} ~, \\
&\hspace{0.75cm}\text{ where }\hspace{0.75cm}
W_{\mathcal{P}, A}=\sum_{s\in\mathcal{P}, i\in A} W_{si} ~.  \nonumber
\end{align}
The sums here are extended over all the spectral pixels belonging to
plate $\mathcal{P}$, and over all the quasars that are at a separation
from pixel $i$ that is within bin $A$. Note that often a bin $i$ in
$\mathcal{P}$ may not appear in the sum if there is no quasar within
bin $A$, but it may also appear repeatedly if more than one quasar is
at a separation within bin $A$. The quasars may be in a plate different
from $\mathcal{P}$ because the angular separation from the CIV spectrum
may take it outside the plate. In practice, this can be calculated by
looping over all pairs of spectra and quasars, and adding the terms in
equation \ref{eq:xi_sub} to the corresponding bins $A$ for each pixel
$i$. The final cross-correlation is obtained as
\begin{align}\label{eq:xi_full}
&\xi_{A}=\sum_{\mathcal{P}} W_{\mathcal{P},A} \xi_{\mathcal{P},A} / W_{A} ~,\\
& \hspace{0.75cm}\text{ where }\hspace{0.75cm}
 W_{A}=\sum_{\mathcal{P}} W_{\mathcal{P},A} ~.  \nonumber
\end{align}

\subsection{The covariance matrix}\label{ssec:covmat}

~\par The covariance matrix of the values of the cross-correlation in
any two bins $A$ and $B$ is
\begin{equation}\label{eq:cov}
 C_{AB} = \left\langle \xi_A\xi_B \right\rangle - 
   \left\langle \xi_A\right\rangle\left\langle \xi_B \right\rangle ~.
\end{equation}

~\par We can obtain this covariance matrix from the values of the
cross-correlation obtained in many independent sub-samples. If we
neglect the small correlations among neighboring sub-samples due to the
correlated large-scale structure, the covariance can be expressed as
\begin{equation}
C_{AB} = \frac{1}{W_AW_B} \,
 \sum_{\mathcal{P}} W_{\mathcal{P},A}W_{\mathcal{P},B}
 \left[\xi_{\mathcal{P},A}\xi_{\mathcal{P},B} - \xi_A\xi_B\right] ~.
\end{equation}
However, for this covariance matrix to be reliably obtained in this way,
its total number of elements needs to be considerably smaller than the
number of sub-samples, $N_p=2370$. The covariance matrix is actually
extremely large (it is a symmetric $512\times 512$ matrix), so it cannot
be computed in this way directly. Instead, we follow the procedure
used and described in \cite{Delubac2015,Bautista2017,Helion2017,Perez2018}.  
We start by defining the normalized covariance
matrix, referred to as the correlation matrix : 
\begin{equation}\label{eq:corrmatrix}
\rho_{AB}=\frac{C_{AB}}{\sqrt{C_{AA}C_{BB}}} ~.
\end{equation}
To a good approximation, we find that $\rho_{AB}$ is a function only of
$\Delta r_{\perp} = |r_{\perp, A} - r_{\perp,B}|$ and
$\Delta r_{\parallel} = |r_{\parallel, A} - r_{\parallel,B}|$. We
average this correlation matrix over all bin pairs having the same
$\Delta r_\perp$ and $\Delta$ \rpar. This averaged correlation matrix
has only $16\times 32=512$ elements, which are sufficiently well
determined from our 2,370 sub-samples. Finally, we re-compute a smoothed
covariance matrix $C_{AB}$ from equation \ref{eq:corrmatrix}, using the
averaged correlation matrix and the original values for the diagonal
elements $C_{AA}$ and $C_{BB}$.

\begin{figure}
\includegraphics[width=0.45\textwidth]{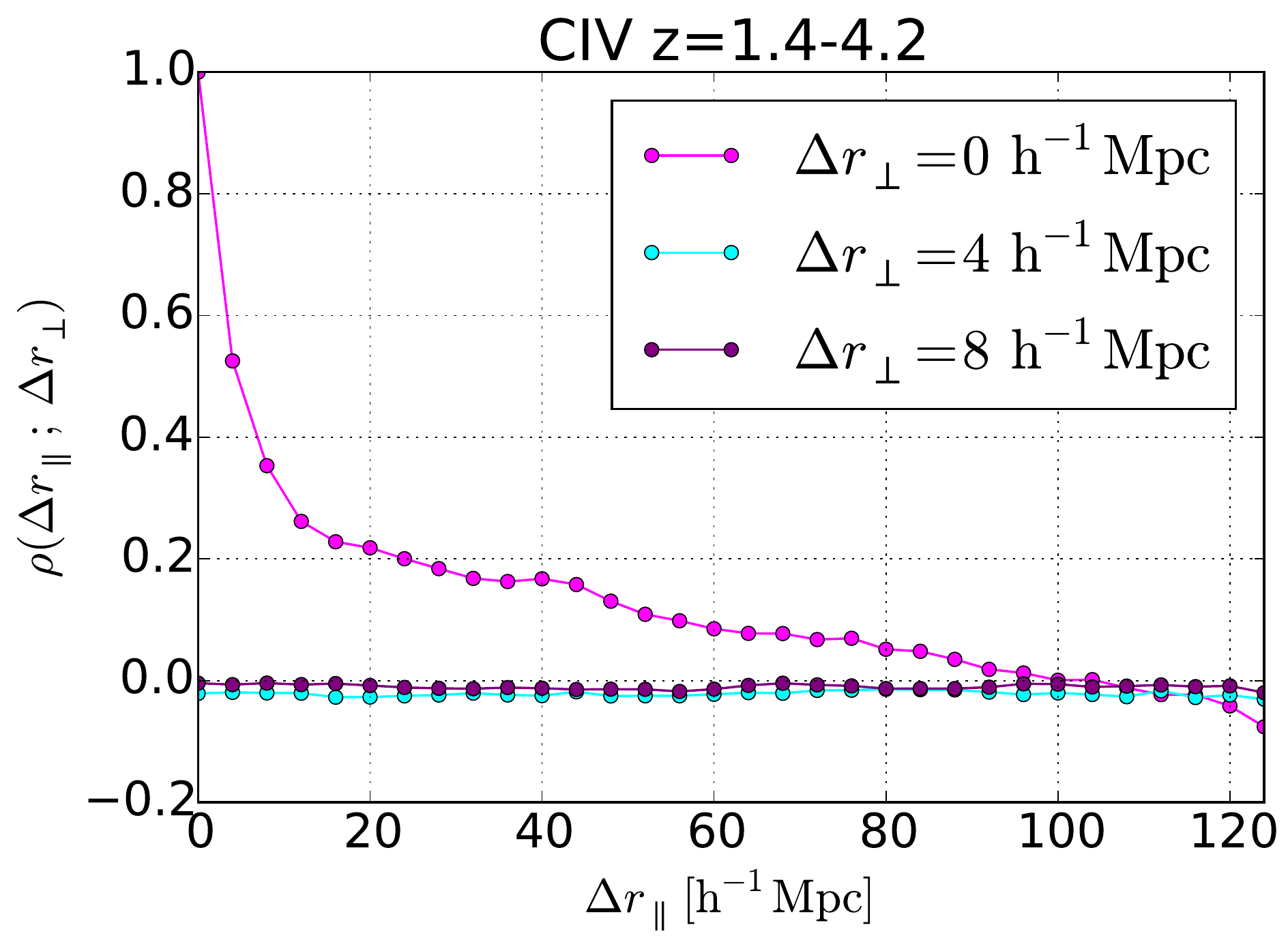}
\caption[\textit{Smoothed} correlation matrix.]
{The averaged correlation $\rho_{AB}=C_{AB}/\sqrt{C_{AA}C_{BB}}$ as a
function of $\Delta r_{\parallel} = |r_{\parallel, A} - r_{\parallel,B}|$,
shown for perpendicular separations
$\Delta r_{\perp} = |r_{\perp, A} - r_{\perp,B}| =0$ (magenta),
$\Delta r_{\perp} = 4 \hmpc$ (cyan) and
$\Delta r_{\perp} = 8 \hmpc$ (purple).
}
\label{fig:corrcoef}
\end{figure} 

~\par Values of the correlation matrix are shown in Figure
\ref{fig:corrcoef} when averaged over our full redshift interval, which
shows that the most important correlation coefficients are for
$\Delta r_\perp=0$. The correlation is primarily due to pairs of
pixel-quasar pairs sharing the same quasar and the same spectrum, which
appear at $\Delta r_\perp =0$ (see section 4.3 and Appendix A.2 of
\cite{Helion2017}).

\subsection{Fitting the CIV bias}
\label{ssec:model}

~\par In the limit of large scales, we can model the cross-correlation
of any two tracers of the large-scale density field using linear theory
\citep{Kaiser1987}. Even though the scale at which we clearly detect
the cross-correlation of quasars and CIV absorption reaches out to only
$\sim 10 \hmpc$, where linear theory may not be very accurate even at
the high-redshift we make our measurements, it still provides a good
model when combined with a velocity dispersion that can account for
non-linear peculiar velocities in addition to quasar redshift errors
and the intrinsic profile of the CIV doublet line. Following the same
formalism as in \cite{Font2012}, the linear cross-correlation between
a transmission fluctuation $\delta_c$ and a set of objects like our
quasar sample can be expressed in terms of its Fourier transform, or
cross-power spectrum,
\begin{align}\label{eq:ps}
& P_{Qc}(\textbf{k},z) =\\
&b_{Fc}(z)\, [1+\beta_{c}(z)\, \mu_k^2] \cdot
  b_q(z)[1+\beta_q(z)\, \mu_k^2]\, P_L(k,z) ~,  \nonumber
\end{align}
where $b_{Fc}$ is the bias factor of the CIV transmission
fluctuation, $b_q$ is the quasar bias factor, and $\beta_{c}$ and
$\beta_q$ are the redshift space distortion parameters for CIV and for quasars. 
The bias $b_{Fc}$ is not the usual bias factor describing the large-scale 
distribution of a population of objects, but is 
similar to the \lya forest bias factor that relates a transmission
fluctuation to a linear mass density fluctuation on large scales
(see \citep{Slosar2011,FontMiralda2012}): it has a negative
value because a mass overdensity results in a reduced transmission, and
a very small absolute value that reflects the small average absorption of
the CIV forest and the corresponding small fluctuation in the transmission.
This will be clarified in section \ref{sec:ccl}, where the relation of
$b_{Fc}$ to the bias factor of the population of CIV absorbers will be
discussed. The linear matter power spectrum is $P_L(k,z)$; as expressed
in equation (\ref{eq:ps}), the amplitude of each Fourier mode in a
biased tracer field is enhanced by the redshift distortion factor
$b_{Fc}(z)[1+\beta_{c}(z)\, \mu_k^2]$ of the CIV absorption, and by the
corresponding factor for the quasars, where $\mu_k$ is the cosine of the
angle between $\vec{k}$ and the line of sight.

~\par We complement this linear theory formula by multiplying $P_{Qc}$ by
a Gaussian in the parallel direction, with a dispersion $\sigma_\parallel^{-1}$ 
(where the cross-correlation is convolved with a Gaussian with dispersion
$\sigma_\parallel$), and adding a shift $\Delta v$ in the parallel direction to
account for a possible systematic in the quasar redshift error that
displaces the center of the cross-correlation. The dispersion can
account for non-linearities, peculiar velocity dispersions of the quasar
and CIV clouds, quasar redshift errors, and the mean profile of the CIV
doublet line. The convolution of all these functions is approximated as
a Gaussian in our analysis. We ignore any effect of the continuum fitting,
which is less important than for the \lya forest (e.g., \cite{Bautista2017}
because of the low mean absorption by metal lines and the fact that we
do not use the flux in the CIV forest interval itself to determine the
continuum.

~\par We make use of the fitting tool Baofit 
\footnote{\href{http://darkmatter.ps.uci.edu/wiki/DeepZot/Baofit}{http://darkmatter.ps.uci.edu/wiki/DeepZot/Baofit}}
developed in the context of the BOSS collaboration
\citep{Kirkby2013,Blomqvist2015} to compute parameterized fits
with equation (\ref{eq:ps}) of correlations of any tracer populations.

~\par The signal-to-noise of our detection of the CIV-quasar cross-correlation is
not high enough to simultaneously fit a large number of parameters.
We fix the quasar bias and redshift distortion factors to values determined
from other observations, and we vary four free parameters: the CIV
transmission bias and redshift distortion parameters, which are assumed
constant in redshift, and the mean shift and dispersion in the parallel
direction. The mean shift is assumed constant with redshift as a 
velocity, while the dispersion is assumed constant in comoving spatial units.

~\par The fit is done over a range in $r=(r_\parallel^2+r_\perp^2)^{1/2}$ from
$5 \hmpc$ to $60 \hmpc$. Any bins centered at values of $r$ outside this
range are excluded from the fit. This is done to exclude the central
values which are more strongly affected by non-linear effects and the
CIV doublet line shape, and to have a more isotropic distribution of
bins.

~\par We assume that there is no redshift evolution for the CIV bias and 
that the quasar bias follows the power-law evolution
\begin{equation} 
b_{q}(z) = b_{q}(z_{\rm ref}) 
    \left(\frac{1+z}{1+z_{\rm ref}}\right)^{\gamma_q} ~,
\end{equation}
with $b_q(z_{\rm ref})=3.91$ at $z_{\rm ref} = 2.39$,
and $\gamma_q=1.7133$. This power-law was fitted to the recent 
measurements of quasar clustering presented in \cite{Laurent2016}.


\section{Results}\label{sec:results}

~\par In this section the results of the redshift space
cross-correlation of the CIV absorption with quasars are presented.
After obtaining fits to our fiducial model, we check the consistency of
our estimated errors from the smoothed covariance matrix and the
bootstrap method. 

\subsection{The Quasar -- CIV forest cross-correlation}
\label{ssec:qccross}

\begin{figure}
\includegraphics[width=0.48\textwidth]{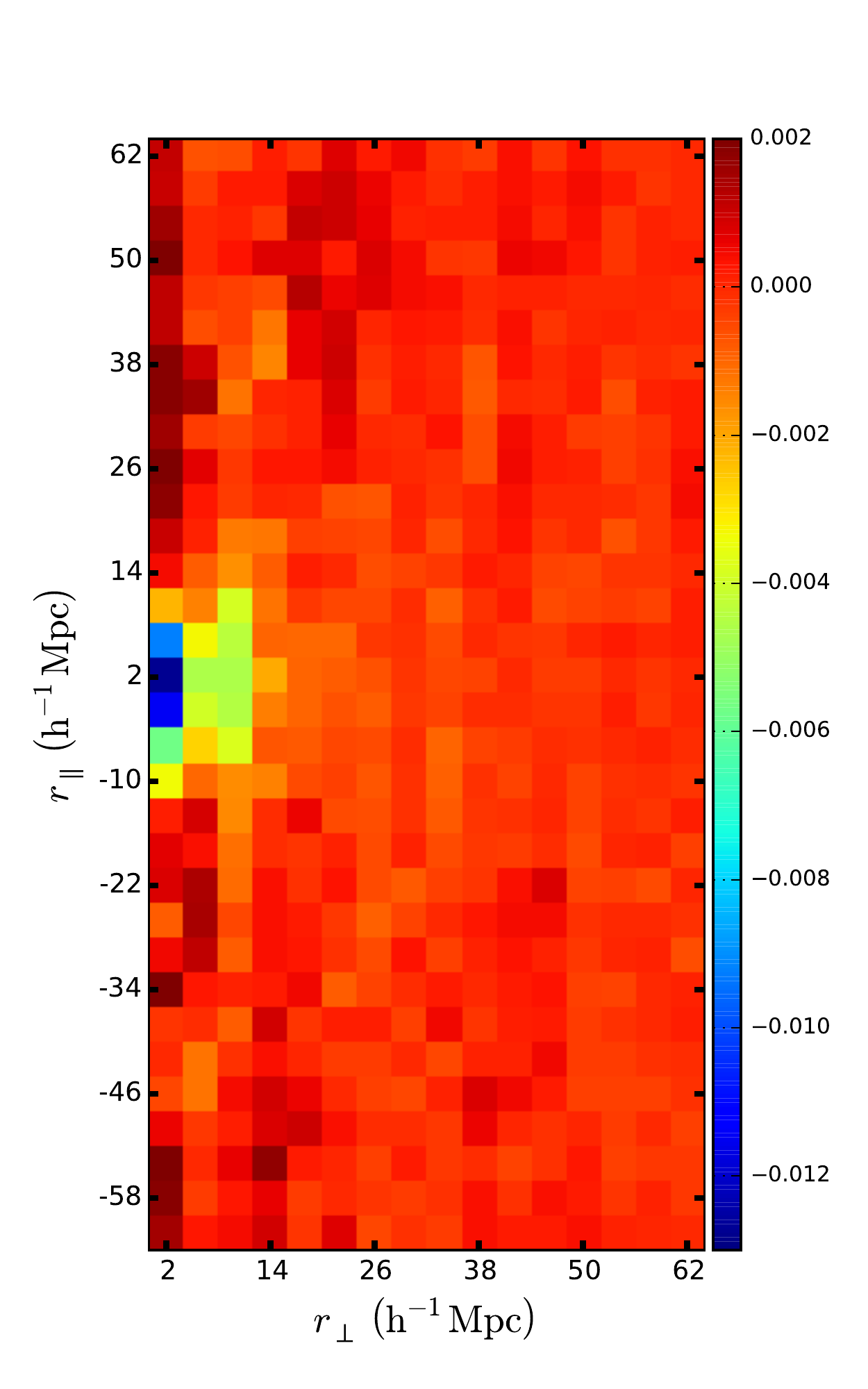}
\caption{Contour plot of the quasar-CIV cross-correlation for the
$\mathbb{F}$ sample covering the full redshift range, as a function of 
transverse separation $r_\perp$ and parallel separation $r_\parallel$.
}
\label{fig:2DFULL}
\end{figure}

~\par Figure \ref{fig:2DFULL} shows this
cross-correlation for the full redshift interval
($\mathbb{F}$ sample). The cross-correlation is in general negative
because the overdensity near the quasar induces a negative fluctuation
of the transmission fraction $\delta_c$ (or a positive fluctuation of
the absorbed fraction by CIV systems).
There is a clear detection at small separations, and a strong radial
elongation at $r < 8 \hmpc$ which is expected from non-linear
peculiar velocities, redshift errors and the CIV doublet absorption
line. The linear redshift space distortion causing a tangential
elongation is not clearly detected. 

\begin{figure*}
\includegraphics[width=1.\textwidth]{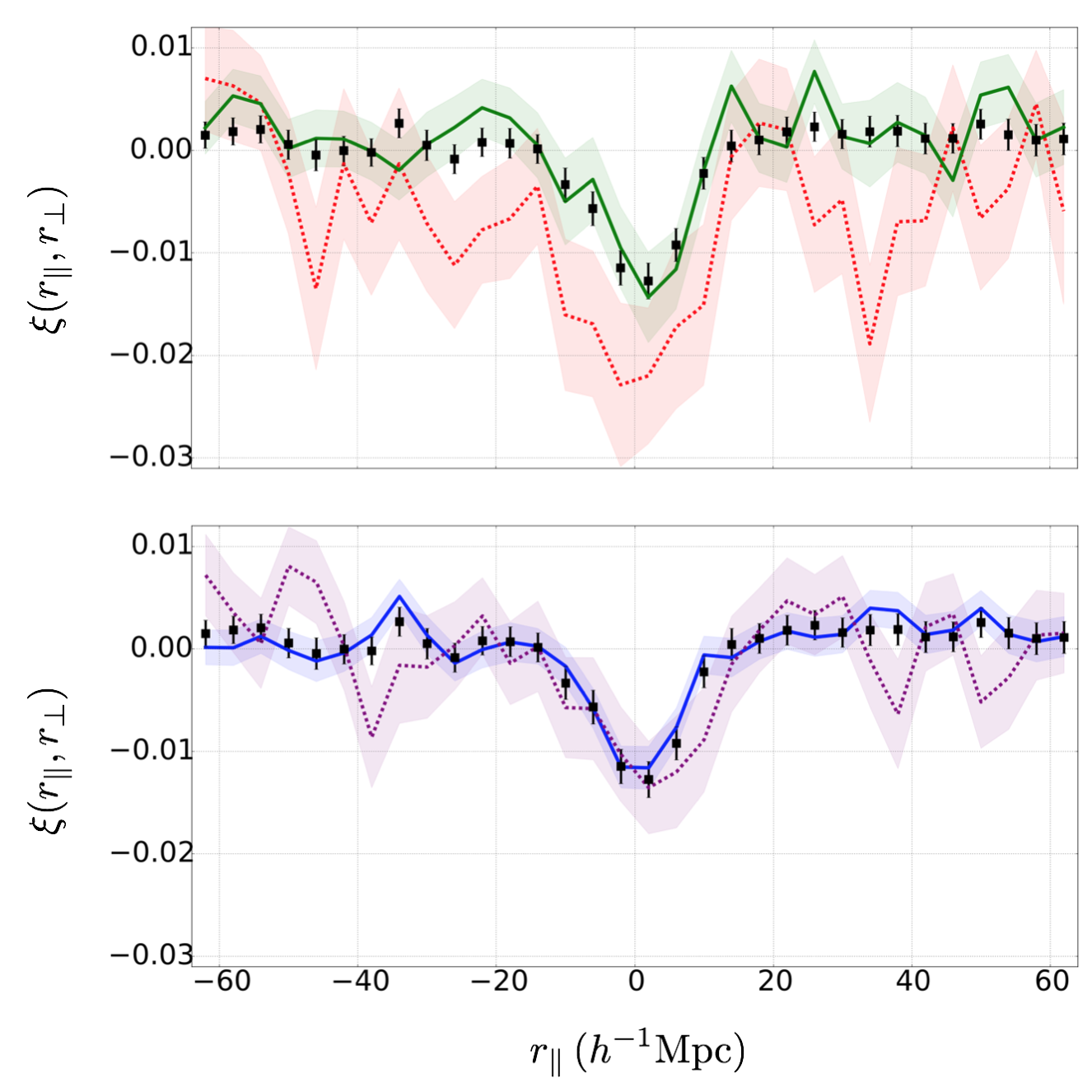}
\caption{Comparison of the quasar-CIV cross-correlation at the
$0 < r_\perp < 4\hmpc$ bins, as a function of \rpar, for different redshift 
samples (previously described in Table \ref{tab:z}). 
Respectively between :  the $\mathbb{F}$ sample (black points with errorbars, for full redshift interval), on the upper panel the v$\mathbb{L}$
sample in red (dotted colored line with shaded area indicating errors, for the lowest
redshift interval) and the $\mathbb{L}$ sample in green solid line (green shaded area for the errors ) 
and on the lower panel the $\mathbb{M}$ sample in blue solid line (blue shaded area for the errors)
and the $\mathbb{H}$ sample in purple (dotted colored line with shaded area indicating errors, for the highest
redshift interval).}
\label{fig:1D}
\end{figure*}

 ~\par  We can better visualize the form of the cross-correlation in the
radial direction in \ref{fig:1D}  
which shows the values measured in
the bins $0.0 < r_{\perp} < 4.0 \hmpc $, as a function of \rpar, for the 
different redshift samples. 
The black squares with error bars are the same in every plot and they show
results for the $\mathbb{F}$ sample, to be compared with the same
measurements at every redshift interval. The four redshift intervals are
shown in the four separate plots as colored lines, and the shaded areas
indicate $1-\sigma$ error bars. There is no clear evidence for any
variation of the cross-correlation with redshift. Only at the lowest
redshift interval (the v$\mathbb{L}$ sample) there is an indication of
a stronger cross-correlation amplitude, but the difference is not highly
significant. Note that the more precise determination of $\xi$ is for
the $\mathbb{M}$ redshift interval, because it contains more than half
of all the quasar-pixel pairs, and has therefore strongly correlated
results with the $\mathbb{F}$ sample. The lowest number of pairs occurs
for the v$\mathbb{L}$ sample, which has the largest errors.

\begin{figure*}
\includegraphics[width=1.\textwidth]{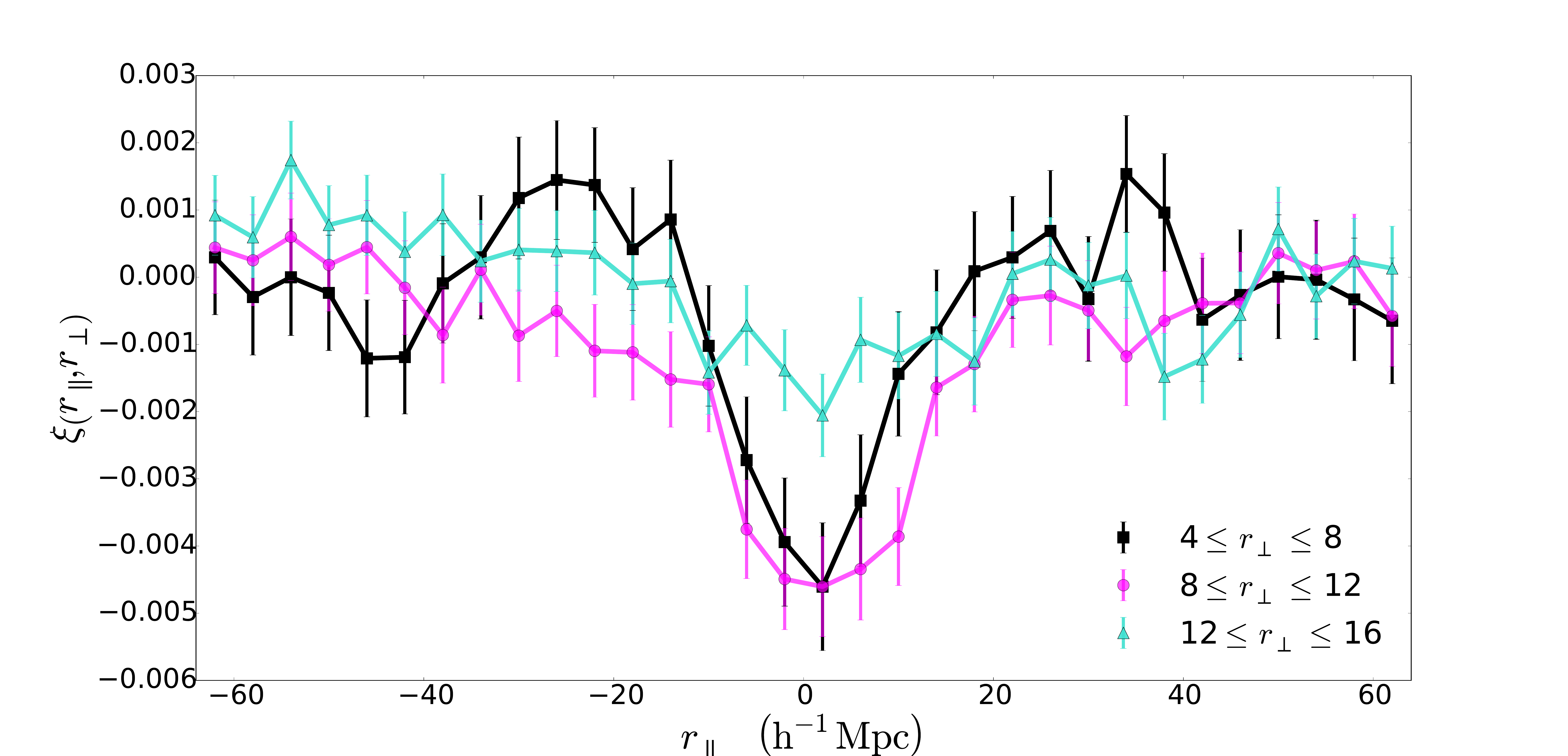}
\caption{Cross-correlation for the $\mathbb{F}$ sample for the indicated 
perpendicular separation $r_\perp$ bins (in units of comoving $\hmpc$), 
as a function of line of sight separation $r_\parallel$.
}
\label{fig:1DFull}
\end{figure*}

~\par Finally, in Figure \ref{fig:1DFull} the cross-correlation is shown
for the $\mathbb{F}$ sample in three bins of perpendicular separation,
from 4 to $16 \hmpc$. The amplitude of the cross-correlation variation
declines with $r_\perp$ as expected. 

\subsection{Fits to the CIV transmission bias parameter}
\label{ssec:cbias}

\begin{table*}
\begin{tabular}{ccccccc}
\hline
  &$(1+\beta_{c})b_{Fc}$ & $\beta_{c}$ & $\Delta_v$ [${\rm km\, s}^{-1}$] &
 $\sigma_{\parallel}$ $[h^{-1}\, {\rm Mpc}]$ & $\chi^2$ & dof \\
\hline
\hline
$\mathbb{F}$ & $-0.024 \pm 0.003$ & $1.09 \pm 0.56$ & $-119.7 \pm 52.4$ & $4.75 \pm 0.83$ & $351.092$ & $356 - 4$ \\
\hline
\hline
$\mathbb{F}$ & $ -0.0209 \pm 0.0015 $ & $ 0.5 $ & $-118.7 \pm 50.7$ & $4.15 \pm 0.71$ & $352.751$ & $356 - 3$ \\
$\mathbb{F}$ & $ -0.0221 \pm 0.0016 $ & $ 0.7 $ & $-119.0 \pm 51.3$ & $4.39 \pm 0.70$ & $351.719$ & $356 - 3$ \\
$\mathbb{F}$ & $ -0.0236 \pm 0.0018 $ & $ 1.0 $ & $-119.5 \pm 52.1$ & $4.68 \pm 0.70$ & $351.12$ & $356 - 3$ \\
\hline
v$\mathbb{L}$ & $-0.0241 \pm 0.0065$ & $1.09$ & $-119.7$ & $4.75$ & $371$ & $356 - 1$ \\ 
$\mathbb{L}$ & $-0.0222 \pm 0.0038$ & $1.09$ & $-119.7$ & $4.75$ & $345$ & $356 - 1$ \\ 
$\mathbb{M}$ & $-0.0234 \pm 0.002$ & $1.09$ & $-119.7$ & $4.75$ & $365$ & $356 - 1$ \\ 
$\mathbb{H}$ & $-0.0246 \pm 0.0054$ & $1.09$ & $-119.7$ & $4.75$ & $342$ & $356 - 1$ \\ 
\hline
\end{tabular}
\caption{Results of the fit parameters $(1+\beta_{c})b_{Fc}$,
$\beta_{c}$, $\Delta_v$ and $\sigma_\parallel$. The first row is for all
4 parameters left free. In the next three rows, we fix the parameter
$\beta_c$ to the values indicated and leave free the other three
parameters. The bottom 4 rows are 1-parameter fits to each
of the 4 redshift sub-samples, varying only $b_{Fc}$ and keeping the
other 3 parameters to the same values as in the first row. \jordi{
If the other parameters were allowed to vary, the marginalized
error bars of the bias would be larger, but the quoted errors
indicate the uncertainty on the bias evolution if we assume the other
parameters do not vary significantly. }}
\label{tab:bias_results}
\end{table*}

~\par We obtain several fits of the quasar-CIV cross-correlation to the
model described in section \ref{ssec:model}, in the redshift intervals
defined in table \ref{tab:z}. 
The parameters we fit are the combination $(1+\beta_{c})b_{Fc}$, where 
$b_{Fc}$ is the CIV bias, the CIV redshift distortion parameter $\beta_{c}$, 
the velocity shift $\Delta_v$ and the dispersion in quasar redshift
$\sigma_{\parallel}$. We start with a fit with all 4-parameters allowed
to vary for the full redshift interval $\mathbb{F}$, with the result
shown in the first row of table \ref{tab:bias_results}. We find that
$\beta_c$ has a large error because of the high degeneracy with
$\sigma_\parallel$, which is broken only by measurements at large $r$
with a low signal-to-noise ratio.

~\par  We therefore present fits with $\beta_c$ fixed to three different values,
to see the variation of the other parameters when $\beta_c$ changes.
As discussed below, the redshift distortion parameter is related to the
bias factor of the CIV absorbers, $b_{\tau c}$, by
$\beta_c=f(\Omega)/b_{\tau c}$, where $f(\Omega)$ is logarithmic
derivative of the growth factor that appears in linear theory
\cite{Kaiser1987}. A value $\beta_c=f(\Omega)\simeq 1$ corresponds to the
case where CIV absorbers have the same large-scale fluctuation amplitude
as the mass density perturbations, while $\beta_c=0.5$ is expected if
CIV absorbers are biased in the same way as DLAs \cite{Perez2018}.
The values of $\chi^2$ indicate that all these fits are consistent with
the data within the uncertainties. The shift parameter $\Delta_v$
remains practically constant, while $\sigma_\parallel$ increases as $\beta_c$
is increased. This correlation of the fit parameters occurs because
increasing $\beta_c$ makes the contours of the cross-correlation more
tangentially elongated, while increasing $\sigma_\parallel$ makes them more
radially elongated.

~\par   Then we fix $\beta_{c}$, $\Delta_v$ and
$\sigma_{\parallel}$ to the values measured in the first fit, and we
do a 1-parameter fit in each of the four redshift sub-samples. These
results are presented in the bottom 4 rows of table
\ref{tab:bias_results}. There is no evidence for any redshift variation
of the bias factor $b_{Fc}$. This transmission bias factor stays
constant within $\sim 20\%$ over our entire redshift range.

\subsection{Bootstrap evaluation of the error}

~\par To verify the robustness of the smoothing of the covariance matrix
described in section \ref{ssec:covmat}, we estimate the error on the 
combination $(1+\beta_{c})b_{Fc}$ by bootstrap analysis with the 2,370 plate
measurements of the cross-correlation. We create 250 bootstrap samples
by picking at random, with repetitions, 2,370 $\xi_{\mathcal{P},A}$
cross-correlations from individual plates, and adding them accounting
for their weights $W_{\mathcal{P},A}$. Then the CIV bias is fitted again
for each of the bootstrap samples, and we obtain a final value and error
for $b_{Fc}$ from the mean and dispersion of these 250 bootstrap values.
We find that this bootstrap error of the CIV bias matches the one
obtained using the smoothed covariance matrix within 3 \%.

\subsection{Bias parameter with the smallest error}

~\par   In previous papers on measuring the bias factor of absorbers, the
combination $b_F(1+\beta)$ has usually been used as one of the
parameters to fit together with $\beta$, because of the small
correlation of errors of these two quantities \cite{Slosar2011}.
Here, however, we can more precisely determine the combination that
stays constant as we vary $\beta_c$, and has the smallest relative
error marginalized over the other parameters. From table
\ref{tab:bias_results}, we find this combination to be
(with a relative error that is the same as for the values listed in
table \ref{tab:bias_results} when $\beta_c$ is fixed),
\begin{equation}\label{eq:bresb}
b_{Fc}(1+0.44\times\beta_c) = -0.0170 \pm 0.0013 ~.
\end{equation}
This is a useful way to express our result, as the parameter we can
measure with the highest accuracy.


\section{On the transmission bias}\label{sec:tbias}

~\par This paper has focused on deriving the transmission bias factor of
the CIV absorption from the measurement of the cross-correlation of this
absorption with quasars. However, a theoretical interpretation of this
measurement requires a relation between this transmission bias factor and
the large-scale bias factor of CIV absorbers. The bias of the absorbers
is the ratio of the relative fluctuation in the density of absorbers to
the relative mass fluctuation, and is equal to the large-scale bias
factor of their host halos.

 ~\par We assume that this relation between the transmission and
absorber bias factors is the general one derived in
\cite{FontMiralda2012} for any class of absorption systems with
absorption profiles that rarely overlap and that are correlated only
with their host halo properties, but not with the surrounding
large-scale structure (see their section 4.3).
We summarise here the derivation of this relation, starting with 
the definition of the transmission fluctuation of CIV
absorbers $\delta_{c}$, $F_c=\overline F_c (1+\delta_{c})$, where $F_c$ is the
flux transmission fraction due to CIV absorbers, and $\overline F_c$ is
its average value. Defining the effective optical depth as
$F_c=e^{-\tau_c}$ when averaged over a large-scale region, our
assumption is that the effective optical depth is the quantity that
changes linearly with the density of absorbers in redshift space, and
therefore the transmission fluctuation $\delta_{c}$ corresponds to an
absorber density fluctuation $\delta_{\tau c}$ that is derived as:
\begin{align}
 \overline F_c (1+\delta_{c}) &= e^{-\tau_c} =e^{-\bar\tau_c
 (1+\delta_{\tau c})} , \\ 
 & =
 \overline F_c (1-\bar\tau_c \delta_{\tau c}) ~ ;   \nonumber \\
 & \Longrightarrow \delta_{\tau c} =  \frac{\delta_c}{\log \overline F_c}~. \nonumber 
\end{align}
The transmission and absorber bias factors and redshift distortion
factors are therefore related as
\begin{equation}
 b_{\tau c}= {b_{Fc} \over \log\overline F_c} ~; \qquad
 \beta_{\tau c} = \beta_{c} ~.
\end{equation}
This absorber bias factor should be equal to the bias factor of their
host halos. Note, however, that the host halo bias may not be just a
function of the halo mass, if other properties such as assembly bias are
important and correlate with the CIV cross section.

 ~\par  In general, the redshift distortion factor should be given by the
equation $\beta_{c}= b_{\tau \eta c} f(\Omega)/b_{\tau c}$, where
$f(\Omega)$ is the logarithmic derivative of the growth factor, and
$b_{\tau\eta c}$ is the peculiar velocity gradient bias of the
absorbers. For this discussion, we assume that this peculiar velocity
gradient bias is unity, and therefore
$\beta_{c}= f(\Omega)/b_{\tau c}$, which is correct if the
absorption profiles depend only on the host halo internal
dynamics and are independent of the large-scale structure around
them. This assumption is not true for the \lya forest, where the
absorption profiles are frequently overlapping and depend on the
large-scale peculiar velocity gradient (see \citet{Arinyo2015}).

 ~\par Our measurement of $\beta_{c}$ is therefore directly related to
the absorber bias, but its measurement error is very
large. To obtain a more reliable estimate of the absorber bias factor
from our measurement of $b_{Fc}(1+\beta_{c})$, we need an independent
estimate of $\bar \tau_c$.

\subsection{The mean transmission of the CIV forest}

 ~\par For a population of uncorrelated absorbers, the mean transmission
is related to the density of absorbers per unit redshift and equivalent
width $W_c$, which we denote as ${\cal{N}} (W_c,z)\, dW_c$. For the CIV
doublet, the standard convention used in the literature is that the
equivalent width $W_c$ is
that of the strongest line at $\lambda = 1548.2$ \AA. The mean
effective optical depth due to the CIV forest is then
\begin{align}
\label{eq:fbint}
 \bar\tau_c(z)&= - \log(\overline F_c(z))\\ &= \int_0^\infty dW_c\,
 {{\cal{N}}(W_c,z)\, W_c (1+\bar q_c) (1+z)
 \over \lambda_c } ~,  \nonumber
\end{align}
where $\lambda_c=1549.1$ \AA \, is the mean rest-frame wavelength of the
CIV doublet, and $\bar q_c$ is the average doublet equivalent width
ratio of the absorbers (weighted by $W_c$).

  ~\par  To estimate $\bar\tau_c$, we use the observed equivalent width
distributions obtained by \cite{Dodorico2010} and \cite{Cooksey2013}. 
The results of \cite{Cooksey2013} are more reliable at high
equivalent width because they are based on a very large sample of
quasars (SDSS DR7), whereas \cite{Dodorico2010} use a smaller sample of
quasar spectra with high resolution and signal-to-noise, allowing them
to measure the distribution of weak absorption systems. We use an
exponential distribution, which fits well the observations of
\cite{Cooksey2013} :
\begin{equation}
{\cal{N}}(W_c,z) = k \exp (- W_c/W_*) ~.
\end{equation}
We take the fitted values for $k$ and $W_*$ from Table 4 of
\cite{Cooksey2013}, for their redshift range $2.24 \leq z < 2.51$, which
best corresponds to our middle redshift range $\mathbb{M}$
($2.15 \leq z < 2.6$) containing most of our absorption systems.
The values are
$W_* = 0.368$ \AA , and $kW_* = 4.84$ (we have converted the value
of $k$ reported by \citet{Cooksey2013} to a density of systems per unit
redshift, which is the directly observed quantity, instead of density
per unit $X$, where $dX=dz (1+z)^2 H_0/H(z)$, a quantity often used
in studies of absorption systems). The distribution of CIV
equivalent widths has been found to evolve only weakly over the
redshift range covered by the SDSS data. We ignore any possible
redshift evolution and we use our results for the full sample
(see their Table 4).

  ~\par Integrating equation (\ref{eq:fbint}) with the exponential form of
the equivalent width distribution, we find
\begin{equation}
 \bar\tau_c= { (1+\bar q)(1+z) k W_*^2 \over \lambda_c} ~.
\end{equation}
We use a mean doublet ratio $\bar q=0.695\pm 0.010$ from the mean value
measured in \cite{MasRibas2016} for CIV systems associated with DLAs,
noting that the mean equivalent width of the CIV line in DLAs found in
\cite{MasRibas2016} is $\bar W_c=0.429$ \AA , only slightly larger than
$W_*$, implying a similar degree of line saturation and therefore of the
value of $\bar q$. At a mean redshift $z=2.37$, this yields a value
$\bar\tau_c=0.00657$.

 ~\par We now correct this value for the fact that the true equivalent
width distribution does not follow the exponential form at $W_c < 0.5$
\AA, where the SDSS data used in \cite{Cooksey2013} starts being
incomplete. Instead, the CIV equivalent width distribution can be fitted
by the power-law ${\cal{N}}(W_c,z) \propto W_c^{-\alpha}$ at low
equivalent widths, where $\alpha=1.53$ fits well the results of
\cite{Dodorico2010} down to the lowest equivalent widths at which they
are complete (see their Figure 2). We assume we
can treat the systems as optically thin for the low column densities
at which we use their model, so that CIV column densities are
proportional to equivalent widths. We match the power-law and equivalent
width distributions at $W_c=1.4 W_* = 0.515$ \AA, which is roughly where
the results of \cite{Dodorico2010} and \cite{Cooksey2013} match (see Figure 10
in \cite{Cooksey2013} ), and also the equivalent width below which the data
of \cite{Cooksey2013} start being incomplete. Assuming a fixed value
$1+\bar q=1.695$, we replace the function $\cal{N}$ by
\begin{equation}
 {\cal N}_{\text{corr}} = k\, \exp(-W_c/W_*) ~, \  (W_c > 1.4 W_*) ~;
\end{equation}
\begin{align}\label{eq:dOd}
 {\cal N}_{\text{corr}} = k\, \exp&(-1.4)\, \left({W_c\over 1.4 W_*} \right)^{-1.53} ~, \\
 &\quad (W_c < 1.4 W_*) ~.  \nonumber
\end{align}
The corrected value with this distribution is found to be
$\bar\tau_c = 0.0106$. Finally, we note that the correction to this
effective optical depth arising from the difference between the
exponential distribution and the power-law distribution at
$W_c < 1.4 W_*$, which is $\Delta \bar\tau_c = 0.0040$, probably has an
average value of $1+\bar q$ that is intermediate between the value
$1+\bar q =1.695$ for DLAs found by \cite{MasRibas2016} , and the limiting
value for optically thin absorbers, $1+\bar q=1.5$. Adopting an
intermediate value $1+\bar q=1.6$ for this correction to the $W_c$
distribution, we reduce the correction to
$\Delta \bar\tau_c = 0.0040\cdot (1.6/1.695)$, and we obtain our final
estimate $\bar\tau_c = 0.0103$.

   ~\par The error on our estimate of $\bar\tau_c$ is dominated by the
uncertainties in the $W_c$ distribution and its redshift evolution, in
particular by the slope of the power-law distribution and how far down
it extends to low $W_c$. An error $\pm 0.1$ to the power-law slope in
$\alpha$ implies an error of $\sim 10\%$ in $\bar\tau_c$. We note also
that the low-$W_c$ absorbers tend to be clustered with the high $W_c$
ones because they often result from de-blending of complex absorption
profiles, which may have led us to an overestimate of $\bar\tau_c$, but
this is difficult to estimate. We therefore urge the total absorption by
all detected absorbers to be directly reported in future studies of
weak metal-line systems.

 ~\par \changes{An alternative method to measure the mean CIV optical depth is by directly
using the pixel distribution of the ratios of CIV to HI optical depths.
The results obtained with this method by \cite{Schaye2003} show a power-law
relation of the CIV and HI optical depths with the approximate form
$\tau_{c}\propto \tau_{_{HI}}^{0.8}$, down to the lowest detectable values
(see their figure 4). This suggests that the power-law equivalent width
distribution of \cite{Dodorico2010}, with index $-1.53$ (eq. \ref{eq:dOd}),
translates to a power-law equivalent width distribution for \lya forest
absorbers with index $-1.53\times 0.8 +0.2 = 1.42$ similar to what is
observed at low column densities for the \lya forest. A large contribution
to the total CIV optical depth from very weak CIV absorbers would require
either a steepening of the HI column density distribution at low $N_{HI}$,
or a flattening of the relation of $\tau_{c}$ to $\tau_{_{HI}}$, which
the observations do not indicate.}


\section{Discussion}\label{sec:ccl}

\subsection{The bias of the CIV absorbers}
\label{ssec:physb}

  ~\par Having estimated $\bar\tau_c$, we can now infer the
average bias of the CIV absorbers. We use our result for the parameter
with the smallest relative error from equation \ref{eq:bresb},
\begin{equation}
  (1+0.44\beta_c) b_{Fc} \simeq -0.0170 \pm 0.0013 ~.
\end{equation}
Using $\beta_c=f(\Omega)/b_{\tau c}$, $b_{Fc}=-\bar\tau_c b_{\tau c}$, and a
value $f(\Omega) \simeq 0.97$ for the cosmological model we use at the
mean redshift of our sample, we obtain
\begin{equation}
 (0.44 + 1/\beta_c) \bar\tau_c = 0.0175 \pm 0.0014 ~.
\end{equation}
If the mean optical depth is in the range $0.008 < \bar\tau_c < 0.012$
(a generously broad range given our previous discussion
on uncertainties in the determination of this quantity),
the previous equality implies $0.6 < \beta_c < 1.0$. This is consistent
with our measurement of $\beta_c$ from the anisotropy of the quasar-CIV
cross-correlation, and with a bias factor of the CIV absorbers that is
roughly in the range $1 < b_{\tau c} < 1.7$. This bias factor is less than that
of DLAs, for which $b_{\rm DLA}\simeq 2$ \cite{Perez2018}, indicating
that the general population of CIV absorbers tends to be in less massive
halos than DLAs.

~\par   This conclusion needs to be taken as preliminary until the value of
$\bar\tau_c$ and the CIV transmission bias factor are measured more
reliably, with a better control on the systematic errors.

\subsection{Relation to the host halo bias factor}

  ~\par \changes{The bias of metal lines can be modeled theoretically assuming that the gas
causing metal absorption is associated with host halos with a certain mass
distribution. Metals are created in halos if the massive stars in which they
are synthesized are all formed in the virialized regions of halos. Galactic
winds can then eject the metal-enriched gas outside the virialized regions,
into the low-density intergalactic gas surrounding halos. The photoionized
intergalactic medium contains substantial CIV down to densities as low as the
mean density of the universe (see e.g., figure 3 in \cite{Schaye2003}), so
many CIV absorbers may arise far from the virialized regions of halos. This,
however, does not alter the equality of the bias factor of the CIV absorbers
to the bias factor of the halos they originated from, on scales larger than
the size of the galactic winds, or the largest distance that the metal-enriched
wind can traverse.}

~\par  Any large-scale structure model predicts a number density of halos
$n(M)\, dM$ of mass $M$ in a bin $dM$. If a line of sight intercepts
a halo, a metal line can be observed with some equivalent width $W_c$.
A useful concept to define is the ${\it absorption\, volume}$ of the
metal line produced by a halo, equal to
\begin{equation}
 V_c(M) =  \int d^2x {W_c (1+\bar q_c)(1+z)\over \lambda_c}
{c\over H(z)} ~,
\end{equation}
where the integral is done over the projected area of the halo seen
from a specific line of sight, with comoving coordinates $x$ in the
perpendicular directions. The volume $V_c$ is
then obtained in comoving units, and is defined to be an average
over all halos of mass $M$. The mean absorption of CIV systems is
then given by
\begin{equation}
 \bar\tau_c =  \int dM\, n(M)\, V_c(M) ~.
\end{equation}
Large-scale structure models predict the bias $b_h(M)$ of halos of
mass $M$, and the bias factors of CIV systems are related to this by
\begin{align}
 &b_{Fc} = - \int dM n(M)\, b_h(M)\, V_c(M) ~; \\ &
 b_{\tau c} = - {b_{Fc}\over \bar\tau_c} =
 {1\over \bar\tau_c} \int dM n(M)\, b_h(M)\, V_c(M) ~.  \nonumber
\end{align}
This relation assumes that the CIV absorption volume $V_c$ does not
depend on any halo property that can alter the value of the bias factor,
except for the halo mass. In particular, the presence of assembly bias
(the dependence of halo bias on its formation time at a fixed mass; see
\cite{Borzyszkowski2017}) alters this relation if $V_c$ depends also on
the halo formation time.

 ~\par  We point out the distinction between the average bias factor of CIV
absorbers, for which every absorber is weighted equally and the host
halos are weighted by their mean cross section to produce an absorber,
and the average bias factor of the CIV transmission fluctuation, which
we have measured here and should be compared to the host halo bias
weighted by the absorption volume we have defined here. The absorption
volume should increase with halo mass faster than the cross section
(because of the increase of velocity dispersion with halo mass), and
therefore the absorption-volume weighted bias factor should be higher
than the cross-section weighted one. This further supports the
conclusion that the CIV absorbers contributing to the mean CIV
transmission are hosted by halos of lower bias (i.e., less massive
halos) than DLAs.

\subsection{Comparison to other work}\label{ssec:comp}

~\par  \cite{Vikas2013}
also measured the cross-correlation of CIV absorbers with quasars, but
using individually identified CIV systems. They translated an
isotropic measurement of the cross-correlation to a CIV bias factor of
$b_c=2.38\pm 0.62$. The bias inferred for their individual CIV
absorbers should be the same as our absorber bias $b_{\tau c}$, under
the assumption that the weak systems that are not detected in the
catalog used by \cite{Vikas2013} have the same bias factor as the mean
of all the CIV systems weighted by their equivalent width, which
determines our CIV transmission bias factor. As mentioned previously,
the principal uncertainty in comparing the two results is the value
of $\bar\tau_c$. In addition, our analysis has included the redshift
space distortion factor $\beta_c$, whereas \cite{Vikas2013} did only
an isotropic analysis. The two measurements are not inconsistent taking
into account the large errors in both of them, however they suggest that
weak CIV systems may have lower bias than strong ones, which would help
explain why the individually detected CIV systems have a higher bias
than the overall population.

~\par \sgg{Next, we comment on the result of \cite{Blomqvist2018}, who have
simultaneously completed a similar analysis to ours including eBOSS
data. Apart from having a larger data set (especially at low redshift),
there are other differences in the two analyses. One is the different
continuum fitting methods.
\cite{Blomqvist2018} apply the same continuum fitting method used in
other BOSS analyses of the Ly$\alpha$ forest, hence the need for a
distortion matrix in their work (i.e. their section 5.2).
They also use an upgrade from the Baofit code we use called \textit{picca}
\footnote{ \href{https://github.com/igmhub/picca/}{https://github.com/igmhub/picca/}},
although this should not affect the results. 
Their fitting range extends to a larger scale than ours: from 10 $\hmpc$
to 180 $\hmpc$, compared to our range from $5\hmpc$ to $60 \hmpc$.}

~\par \sgg{We focus on comparing their result for the bias and redshift
distortion factors when restricting the rest-frame wavelength range
to $1420$ \AA $\, \leq \lambda_{RF} < 1520$ \AA , the same one we have
used which avoids the contribution from SiIV lines. The mean result of
\cite{Blomqvist2018} for their entire redshift range is
$\beta_c=0.35 \pm 0.21$, and $b_c(1+\beta_c)=-0.019 \pm 0.002$. In
comparison, we obtain a higher value of the redshift distortion factor,
$\beta_c=1.09\pm 0.56$, although the large errors do not make the two
values incompatible. Our value for the bias $b_c$ agrees with theirs:
from our equation \ref{eq:bresb}, if we use their central value
$\beta_c=0.35$, we find that our measurement $b_c(1+0.44\beta_c)=-0.017$
implies $b_c(1+\beta_c)=-0.0199\pm 0.0015$ for $\beta_c=0.35$, in full
agreement with \cite{Blomqvist2018}.}

 ~\par  \sgg{The low central value of $\beta_c$ found by \cite{Blomqvist2018} is
not consistent with our constraint derived from the bounds in
$\bar\tau_c$ from measurements of the incidence rate of CIV absorbers as
a function of their equivalent width in section \ref{ssec:physb},
implying $0.6 < \beta_c < 1$. Taking into account the measurement error,
however, their result can be compatible with this constraint.
It is worth noting that their result for $\beta_c$ is substantially
different when dividing the data set into two redshift intervals. For
low redshift ($z< 2.2$ and a mean effective redshift
$z_{\rm eff}=1.69$), they find $\beta_c=0.05\pm 0.19$, whereas for high
redshift ($z>2.2$ and $z_{\rm eff}=2.41$, closer to our measurement with
$z_{\rm eff}=2.29$) their result is $\beta_c=0.67\pm 0.29$. Their result
at high redshift is therefore fully consistent with ours, and with our
derived constraint from the value of $\bar\tau_c$. The very low value
of $\beta_c$ at low redshift derived by \cite{Blomqvist2018} suggests
that the absorber bias factor $b_{\tau c}$ may be increasing rapidly
with decreasing redshift, perhaps because CIV is being destroyed in
low-mass halos and being formed in massive halos. However, more data at
low redshift that allow reduced error bars and testing of systematics is
necessary before one can draw solid conclusions on the redshift
evolution of the bias and redshift distortion factors.}

\section{Conclusions}
\label{sec:conc}

~\par With the final SDSS-III Data Release DR12, we have measured the
cross-correlation of the CIV forest absorption with quasars.
We found that the simple linear
theory model for this cross-correlation, with the redshift
distortions as predicted by \cite{Kaiser1987}, is fully consistent
with the data, and we have obtained the CIV transmission bias factor
required to match the measured cross-correlation amplitude. Our main
results are:

\begin{itemize}
\item We measure $(1+\beta_{c})b_{Fc} = -0.024 \pm 0.003$, and
$\beta_c = 1.1 \pm 0.6$ at redshift $z=2.3$, from a fit obtained over a
radial range $5 \hmpc < r < 60 \hmpc$. The value of $\beta_c$ is highly
uncertain, but we can determine most accurately the combination
$(1+0.44\beta_{c})b_{Fc} = -0.0170 \pm 0.0013$.\\

\item The CIV transmission bias does not show any detectable redshift
evolution over the range $1.72 < z < 2.85$. \\

\item Using a derived value of $\bar \tau_c(z)\simeq 0.01$ from
measurements of the equivalent width distribution of CIV absorbers in
the literature, with a generous uncertainty of 20\% , we infer a
redshift distortion parameter in the range $0.6 < \beta_c < 1$, and an
absorber bias factor in the range $1 < b_{\tau c} < 1.7$, which is
substantially lower than the bias factor of DLAs. This suggests that the
CIV absorption systems dominating the total CIV mean absorption are
hosted in halos of lower mass than DLAs, at our mean redshift
$z\sim 2.3$.
The measurements of $\beta_c$ by \cite{Blomqvist2018} also
suggest this may be changing at lower redshifts, $z\sim 1.7$, with more
CIV systems being present in more massive halos. Better measurements of
$\bar\tau_c$ and $\beta_c$ at different redshifts are required to test
the validity of these preliminary conclusions.

\end{itemize}

\section*{Acknowledgments}

~\par This work was supported in part by Spanish grants AYA-2012-33938
and AYA-2015-71091-P. We thank the anonymous referee for his/her
comments. 
SGG thanks the Lawrence Berkeley Laboratory and the APC for their hospitality during part of 
the time when this work was being carried out. 
SGG also thanks Daniel Margala from UCI and NASA's Craig Gordon for the technical support they provided.
We thank Kathy Cooksey and Mat Pieri for discussions on different methods
to measure $\bar\tau_c$, and the BOSS \lya working group for their comments and
suggestions.
AFR acknowledges support by an STFC Ernest Rutherford Fellowship, grant 
reference ST/N003853/1. This work was partially enabled by funding from 
the UCL Cosmoparticle Initiative.

~\par Funding for SDSS-III has been provided by the Alfred P. Sloan Foundation, the Participating
Institutions, the National Science Foundation, and the U.S. Department of Energy
Office of Science. The SDSS-III web site is \href{http://www.sdss3.org/}{http://www.sdss3.org/}.

~\par SDSS-III is managed by the Astrophysical Research Consortium for the Participating
Institutions of the SDSS-III Collaboration including the University of Arizona, the Brazilian
Participation Group, Brookhaven National Laboratory, University of Cambridge, Carnegie
Mellon University, University of Florida, the French Participation Group, the German Participation
Group, Harvard University, the Instituto de Astrofisica de Canarias, the Michigan
State/Notre Dame/JINA Participation Group, Johns Hopkins University, Lawrence Berkeley
National Laboratory, Max Planck Institute for Astrophysics, Max Planck Institute for
Extraterrestrial Physics, New Mexico State University, New York University, Ohio State University,
Pennsylvania State University, University of Portsmouth, Princeton University, the 
Spanish Participation Group, University of Tokyo, University of Utah, Vanderbilt University,
University of Virginia, University of Washington, and Yale University.


\bibliography{CarbonIVb} {}

\begin{thebibliography}{}
\makeatletter
\relax
\def\mn@urlcharsother{\let\do\@makeother \do\$\do\&\do\#\do\^\do\_\do\%\do\~}
\def\mn@doi{\begingroup\mn@urlcharsother \@ifnextchar [ {\mn@doi@}
  {\mn@doi@[]}}
\def\mn@doi@[#1]#2{\def\@tempa{#1}\ifx\@tempa\@empty \href
  {http://dx.doi.org/#2} {doi:#2}\else \href {http://dx.doi.org/#2} {#1}\fi
  \endgroup}
\def\mn@eprint#1#2{\mn@eprint@#1:#2::\@nil}
\def\mn@eprint@arXiv#1{\href {http://arxiv.org/abs/#1} {{\tt arXiv:#1}}}
\def\mn@eprint@dblp#1{\href {http://dblp.uni-trier.de/rec/bibtex/#1.xml}
  {dblp:#1}}
\def\mn@eprint@#1:#2:#3:#4\@nil{\def\@tempa {#1}\def\@tempb {#2}\def\@tempc
  {#3}\ifx \@tempc \@empty \let \@tempc \@tempb \let \@tempb \@tempa \fi \ifx
  \@tempb \@empty \def\@tempb {arXiv}\fi \@ifundefined
  {mn@eprint@\@tempb}{\@tempb:\@tempc}{\expandafter \expandafter \csname
  mn@eprint@\@tempb\endcsname \expandafter{\@tempc}}}

\bibitem[\protect\citeauthoryear{Ade et~al.}{Ade et~al.}{2016}]{Planck2015}
Ade P. A.~R.,  et~al., 2016, \mn@doi [A\&A] {10.1051/0004-6361/201525830}, 594,
  A13

\bibitem[\protect\citeauthoryear{{Alam} et~al.,}{{Alam}
  et~al.}{2015}]{DR122015}
{Alam} S.,  et~al., 2015, \mn@doi [ApJS] {10.1088/0067-0049/219/1/12}, \href
  {http://adsabs.harvard.edu/abs/2015ApJS..219...12A} {219, 12}

\bibitem[\protect\citeauthoryear{{Arinyo-i-Prats}, {Miralda-Escud{\'e}}, {Viel}
   \& {Cen}}{{Arinyo-i-Prats} et~al.}{2015}]{Arinyo2015}
{Arinyo-i-Prats} A.,  {Miralda-Escud{\'e}} J.,  {Viel} M.,   {Cen} R.,  2015,
  \mn@doi [JCAP] {10.1088/1475-7516/2015/12/017}, \href
  {http://adsabs.harvard.edu/abs/2015JCAP...12..017A} {12, 017}

\bibitem[\protect\citeauthoryear{{Bahcall} \& {Spitzer}}{{Bahcall} \&
  {Spitzer}}{1969}]{Bahcall1969}
{Bahcall} J.~N.,  {Spitzer} Jr. L.,  1969, \mn@doi [ApJ] {10.1086/180350},
  \href {http://adsabs.harvard.edu/abs/1969ApJ...156L..63B} {156, L63}

\bibitem[\protect\citeauthoryear{{Bautista} et~al.,}{{Bautista}
  et~al.}{2017}]{Bautista2017}
{Bautista} J.~E.,  et~al., 2017, \mn@doi [A\&A] {10.1051/0004-6361/201730533},
  603, A12

\bibitem[\protect\citeauthoryear{{Bird}, {Rubin}, {Suresh}  \&
  {Hernquist}}{{Bird} et~al.}{2016}]{Bird2016}
{Bird} S.,  {Rubin} K.~H.~R.,  {Suresh} J.,   {Hernquist} L.,  2016, \mn@doi
  [MNRAS] {10.1093/mnras/stw1582}, \href
  {http://adsabs.harvard.edu/abs/2016MNRAS.462..307B} {462, 307}

\bibitem[\protect\citeauthoryear{{Blomqvist} et~al.,}{{Blomqvist}
  et~al.}{2015}]{Blomqvist2015}
{Blomqvist} M.,  et~al., 2015, \mn@doi [JCAP] {10.1088/1475-7516/2015/11/034},
  \href {http://adsabs.harvard.edu/abs/2015JCAP...11..034B} {11, 034}

\bibitem[\protect\citeauthoryear{{Blomqvist} et~al.}{{Blomqvist}
  et~al.}{2018}]{Blomqvist2018}
{Blomqvist} M.,  et~al., 2018, {preprint,} (\mn@eprint {arXiv} {1801.01852})

\bibitem[\protect\citeauthoryear{{Boksenberg} \& {Sargent}}{{Boksenberg} \&
  {Sargent}}{2015}]{Boksenberg2015}
{Boksenberg} A.,  {Sargent} W.~L.~W.,  2015, \mn@doi [ApJS]
  {10.1088/0067-0049/218/1/7}, \href
  {http://adsabs.harvard.edu/abs/2015ApJS..218....7B} {218, 7}

\bibitem[\protect\citeauthoryear{{Bolton} et~al.,}{{Bolton}
  et~al.}{2012}]{Bolton2012}
{Bolton} A.~S.,  et~al., 2012, \mn@doi [AJ] {10.1088/0004-6256/144/5/144},
  \href {http://adsabs.harvard.edu/abs/2012AJ....144..144B} {144, 144}

\bibitem[\protect\citeauthoryear{{Borzyszkowski}, {Porciani},
  {Romano-D{\'{\i}}az}  \& {Garaldi}}{{Borzyszkowski}
  et~al.}{2017}]{Borzyszkowski2017}
{Borzyszkowski} M.,  {Porciani} C.,  {Romano-D{\'{\i}}az} E.,   {Garaldi} E.,
  2017, \mn@doi [MNRAS] {10.1093/mnras/stx873}, \href
  {http://adsabs.harvard.edu/abs/2017MNRAS.469..594B} {469, 594}

\bibitem[\protect\citeauthoryear{{Bovy} et~al.,}{{Bovy}
  et~al.}{2011}]{Bovy2011}
{Bovy} J.,  et~al., 2011, \mn@doi [ApJ] {10.1088/0004-637X/729/2/141}, \href
  {http://adsabs.harvard.edu/abs/2011ApJ...729..141B} {729, 141}

\bibitem[\protect\citeauthoryear{{Busca} et~al.,}{{Busca}
  et~al.}{2013}]{Busca2013}
{Busca} N.~G.,  et~al., 2013, \mn@doi [A\&A] {10.1051/0004-6361/201220724},
  \href {http://adsabs.harvard.edu/abs/2013A%26A...552A..96B} {552, A96}

\bibitem[\protect\citeauthoryear{Coil et~al.}{Coil et~al.}{2008}]{Coil2007}
Coil A.~L.,  et~al., 2008, \mn@doi [ApJ] {10.1086/523639}, 672, 153

\bibitem[\protect\citeauthoryear{{Cole} \& {Kaiser}}{{Cole} \&
  {Kaiser}}{1989}]{Cole1989}
{Cole} S.,  {Kaiser} N.,  1989, \mn@doi [MNRAS] {10.1093/mnras/237.4.1127},
  \href {http://adsabs.harvard.edu/abs/1989MNRAS.237.1127C} {237, 1127}

\bibitem[\protect\citeauthoryear{{Cooksey}, {Kao}, {Simcoe}, {O'Meara}  \&
  {Prochaska}}{{Cooksey} et~al.}{2013}]{Cooksey2013}
{Cooksey} K.~L.,  {Kao} M.~M.,  {Simcoe} R.~A.,  {O'Meara} J.~M.,   {Prochaska}
  J.~X.,  2013, \mn@doi [ApJ] {10.1088/0004-637X/763/1/37}, \href
  {http://adsabs.harvard.edu/abs/2013ApJ...763...37C} {763, 37}

\bibitem[\protect\citeauthoryear{{Croft}, {Weinberg}, {Katz}  \&
  {Hernquist}}{{Croft} et~al.}{1998}]{Croft1998}
{Croft} R.~A.~C.,  {Weinberg} D.~H.,  {Katz} N.,   {Hernquist} L.,  1998,
  \mn@doi [ApJ] {10.1086/305289}, \href
  {http://adsabs.harvard.edu/abs/1998ApJ...495...44C} {495, 44}

\bibitem[\protect\citeauthoryear{{Croft}, {Weinberg}, {Pettini}, {Hernquist}
  \& {Katz}}{{Croft} et~al.}{1999}]{Croft1999}
{Croft} R.~A.~C.,  {Weinberg} D.~H.,  {Pettini} M.,  {Hernquist} L.,   {Katz}
  N.,  1999, \mn@doi [ApJ] {10.1086/307438}, \href
  {http://adsabs.harvard.edu/abs/1999ApJ...520....1C} {520, 1}

\bibitem[\protect\citeauthoryear{{Croft}, {Weinberg}, {Bolte}, {Burles},
  {Hernquist}, {Katz}, {Kirkman}  \& {Tytler}}{{Croft}
  et~al.}{2002}]{Croft2002}
{Croft} R.~A.~C.,  {Weinberg} D.~H.,  {Bolte} M.,  {Burles} S.,  {Hernquist}
  L.,  {Katz} N.,  {Kirkman} D.,   {Tytler} D.,  2002, \mn@doi [ApJ]
  {10.1086/344099}, \href {http://adsabs.harvard.edu/abs/2002ApJ...581...20C}
  {581, 20}

\bibitem[\protect\citeauthoryear{Croom et~al.,}{Croom et~al.}{2005}]{Croom2004}
Croom S.~M.,  et~al., 2005, \mn@doi [MNRAS] {10.1111/j.1365-2966.2004.08379.x},
  356, 415

\bibitem[\protect\citeauthoryear{{D'Odorico}, {Calura}, {Cristiani}  \&
  {Viel}}{{D'Odorico} et~al.}{2010}]{Dodorico2010}
{D'Odorico} V.,  {Calura} F.,  {Cristiani} S.,   {Viel} M.,  2010, \mn@doi
  [MNRAS] {10.1111/j.1365-2966.2009.15856.x}, \href
  {http://adsabs.harvard.edu/abs/2010MNRAS.401.2715D} {401, 2715}

\bibitem[\protect\citeauthoryear{D'Odorico et~al.}{D'Odorico
  et~al.}{2016}]{DOdorico2016}
D'Odorico V.,  et~al., 2016, \mn@doi [MNRAS] {10.1093/mnras/stw2161}, 463, 2690

\bibitem[\protect\citeauthoryear{{Dawson} et~al.,}{{Dawson}
  et~al.}{2013}]{Dawson2013}
{Dawson} K.~S.,  et~al., 2013, \mn@doi [AJ] {10.1088/0004-6256/145/1/10}, \href
  {http://adsabs.harvard.edu/abs/2013AJ....145...10D} {145, 10}

\bibitem[\protect\citeauthoryear{{Delubac} et~al.,}{{Delubac}
  et~al.}{2015}]{Delubac2015}
{Delubac} T.,  et~al., 2015, \mn@doi [A\&A] {10.1051/0004-6361/201423969},
  \href {http://adsabs.harvard.edu/abs/2015A%26A...574A..59D} {574, A59}

\bibitem[\protect\citeauthoryear{Eftekharzadeh et~al.,}{Eftekharzadeh
  et~al.}{2015}]{Eftekharzadeh2015}
Eftekharzadeh S.,  et~al., 2015, \mn@doi [MNRAS] {10.1093/mnras/stv1763}, 453,
  2779

\bibitem[\protect\citeauthoryear{{Eisenstein} et~al.,}{{Eisenstein}
  et~al.}{2011}]{Eisenstein2011}
{Eisenstein} D.~J.,  et~al., 2011, \mn@doi [AJ] {10.1088/0004-6256/142/3/72},
  \href {http://adsabs.harvard.edu/abs/2011AJ....142...72E} {142, 72}

\bibitem[\protect\citeauthoryear{{Font-Ribera} \&
  {Miralda-Escud{\'e}}}{{Font-Ribera} \&
  {Miralda-Escud{\'e}}}{2012}]{FontMiralda2012}
{Font-Ribera} A.,  {Miralda-Escud{\'e}} J.,  2012, \mn@doi [JCAP]
  {10.1088/1475-7516/2012/07/028}, \href
  {http://adsabs.harvard.edu/abs/2012JCAP...07..028F} {7, 028}

\bibitem[\protect\citeauthoryear{{Font-Ribera} et~al.,}{{Font-Ribera}
  et~al.}{2012}]{Font2012}
{Font-Ribera} A.,  et~al., 2012, \mn@doi [JCAP]
  {10.1088/1475-7516/2012/11/059}, \href
  {http://adsabs.harvard.edu/abs/2012JCAP...11..059F} {11, 059}

\bibitem[\protect\citeauthoryear{{Font-Ribera} et~al.,}{{Font-Ribera}
  et~al.}{2013}]{Font2013}
{Font-Ribera} A.,  et~al., 2013, \mn@doi [JCAP]
  {10.1088/1475-7516/2013/05/018}, \href
  {http://adsabs.harvard.edu/abs/2013JCAP...05..018F} {5, 018}

\bibitem[\protect\citeauthoryear{{Font-Ribera} et~al.,}{{Font-Ribera}
  et~al.}{2014}]{Font2014}
{Font-Ribera} A.,  et~al., 2014, \mn@doi [JCAP]
  {10.1088/1475-7516/2014/05/027}, \href
  {http://adsabs.harvard.edu/abs/2014JCAP...05..027F} {5, 027}

\bibitem[\protect\citeauthoryear{{Fukugita}, {Ichikawa}, {Gunn}, {Doi},
  {Shimasaku}  \& {Schneider}}{{Fukugita} et~al.}{1996}]{Fukugita1996}
{Fukugita} M.,  {Ichikawa} T.,  {Gunn} J.~E.,  {Doi} M.,  {Shimasaku} K.,
  {Schneider} D.~P.,  1996, \mn@doi [AJ] {10.1086/117915}, \href
  {http://adsabs.harvard.edu/abs/1996AJ....111.1748F} {111, 1748}

\bibitem[\protect\citeauthoryear{{Gunn} et~al.,}{{Gunn}
  et~al.}{1998}]{Gunn1998}
{Gunn} J.~E.,  et~al., 1998, \mn@doi [AJ] {10.1086/300645}, \href
  {http://adsabs.harvard.edu/abs/1998AJ....116.3040G} {116, 3040}

\bibitem[\protect\citeauthoryear{{Gunn} et~al.,}{{Gunn}
  et~al.}{2006}]{Gunn2006}
{Gunn} J.~E.,  et~al., 2006, \mn@doi [AJ] {10.1086/500975}, \href
  {http://adsabs.harvard.edu/abs/2006AJ....131.2332G} {131, 2332}

\bibitem[\protect\citeauthoryear{{Hopkins} \& {Beacom}}{{Hopkins} \&
  {Beacom}}{2006}]{Hopkins2006}
{Hopkins} A.~M.,  {Beacom} J.~F.,  2006, \mn@doi [ApJ] {10.1086/506610}, \href
  {http://adsabs.harvard.edu/abs/2006ApJ...651..142H} {651, 142}

\bibitem[\protect\citeauthoryear{{Kaiser}}{{Kaiser}}{1987}]{Kaiser1987}
{Kaiser} N.,  1987, \mn@doi [MNRAS] {10.1093/mnras/227.1.1}, \href
  {http://adsabs.harvard.edu/abs/1987MNRAS.227....1K} {227, 1}

\bibitem[\protect\citeauthoryear{Kirkby et~al.}{Kirkby
  et~al.}{2013}]{Kirkby2013}
Kirkby D.,  et~al., 2013, \mn@doi [JCAP] {10.1088/1475-7516/2013/03/024}, 1303,
  024

\bibitem[\protect\citeauthoryear{{Kirkpatrick}, {Schlegel}, {Ross}, {Myers},
  {Hennawi}, {Sheldon}, {Schneider}  \& {Weaver}}{{Kirkpatrick}
  et~al.}{2011}]{Kirkpatrick2011}
{Kirkpatrick} J.~A.,  {Schlegel} D.~J.,  {Ross} N.~P.,  {Myers} A.~D.,
  {Hennawi} J.~F.,  {Sheldon} E.~S.,  {Schneider} D.~P.,   {Weaver} B.~A.,
  2011, \mn@doi [ApJ] {10.1088/0004-637X/743/2/125}, \href
  {http://adsabs.harvard.edu/abs/2011ApJ...743..125K} {743, 125}

\bibitem[\protect\citeauthoryear{Laurent et~al.}{Laurent
  et~al.}{2016}]{Laurent2016}
Laurent P.,  et~al., 2016, \mn@doi [JCAP] {10.1088/1475-7516/2016/11/060},
  1611, 060

\bibitem[\protect\citeauthoryear{Laurent et~al.}{Laurent
  et~al.}{2017}]{Laurent2017}
Laurent P.,  et~al., 2017, \mn@doi [JCAP] {10.1088/1475-7516/2017/07/017},
  1707, 017

\bibitem[\protect\citeauthoryear{{Lundgren} et~al.,}{{Lundgren}
  et~al.}{2013}]{Lundgren2013}
{Lundgren} B.,  et~al., 2013, in American Astronomical Society Meeting
  Abstracts \#221. p. 402.05

\bibitem[\protect\citeauthoryear{{Lynds}}{{Lynds}}{1971}]{Lynds1971}
{Lynds} R.,  1971, \mn@doi [ApJ] {10.1086/180695}, \href
  {http://adsabs.harvard.edu/abs/1971ApJ...164L..73L} {164, L73}

\bibitem[\protect\citeauthoryear{{Mas-Ribas} et~al.,}{{Mas-Ribas}
  et~al.}{2017}]{MasRibas2016}
{Mas-Ribas} L.,  et~al., 2017, \mn@doi [ApJ] {10.3847/1538-4357/aa81cf}, \href
  {http://adsabs.harvard.edu/abs/2017ApJ...846....4M} {846, 4}

\bibitem[\protect\citeauthoryear{{McDonald}, {Miralda-Escud{\'e}}, {Rauch},
  {Sargent}, {Barlow}, {Cen}  \& {Ostriker}}{{McDonald}
  et~al.}{2000}]{Mcdonald2000}
{McDonald} P.,  {Miralda-Escud{\'e}} J.,  {Rauch} M.,  {Sargent} W.~L.~W.,
  {Barlow} T.~A.,  {Cen} R.,   {Ostriker} J.~P.,  2000, \mn@doi [ApJ]
  {10.1086/317079}, \href {http://adsabs.harvard.edu/abs/2000ApJ...543....1M}
  {543, 1}

\bibitem[\protect\citeauthoryear{{McDonald} et~al.,}{{McDonald}
  et~al.}{2006}]{Mcdonald2006}
{McDonald} P.,  et~al., 2006, \mn@doi [ApJS] {10.1086/444361}, \href
  {http://adsabs.harvard.edu/abs/2006ApJS..163...80M} {163, 80}

\bibitem[\protect\citeauthoryear{{Miralda-Escud{\'e}}, {Cen}, {Ostriker}  \&
  {Rauch}}{{Miralda-Escud{\'e}} et~al.}{1996}]{Miralda1996}
{Miralda-Escud{\'e}} J.,  {Cen} R.,  {Ostriker} J.~P.,   {Rauch} M.,  1996,
  \mn@doi [ApJ] {10.1086/177992}, \href
  {http://adsabs.harvard.edu/abs/1996ApJ...471..582M} {471, 582}

\bibitem[\protect\citeauthoryear{Myers, Brunner, Nichol, Richards, Schneider
  \& Bahcall}{Myers et~al.}{2007a}]{Myers2006a}
Myers A.~D.,  Brunner R.~J.,  Nichol R.~C.,  Richards G.~T.,  Schneider D.~P.,
   Bahcall N.~A.,  2007a, \mn@doi [ApJ] {10.1086/511519}, 658, 85

\bibitem[\protect\citeauthoryear{Myers, Brunner, Richards, Nichol, Schneider
  \& Bahcall}{Myers et~al.}{2007b}]{Myers2006b}
Myers A.~D.,  Brunner R.~J.,  Richards G.~T.,  Nichol R.~C.,  Schneider D.~P.,
   Bahcall N.~A.,  2007b, \mn@doi [ApJ] {10.1086/511520}, 658, 99

\bibitem[\protect\citeauthoryear{{P{\^a}ris} et~al.,}{{P{\^a}ris}
  et~al.}{2012}]{Paris2012}
{P{\^a}ris} I.,  et~al., 2012, \mn@doi [A\&A] {10.1051/0004-6361/201220142},
  \href {http://adsabs.harvard.edu/abs/2012A%26A...548A..66P} {548, A66}

\bibitem[\protect\citeauthoryear{{P{\^a}ris} et~al.,}{{P{\^a}ris}
  et~al.}{2014}]{Paris2014}
{P{\^a}ris} I.,  et~al., 2014, \mn@doi [A\&A] {10.1051/0004-6361/201322691},
  \href {http://adsabs.harvard.edu/abs/2014A%26A...563A..54P} {563, A54}

\bibitem[\protect\citeauthoryear{P{\^a}ris et~al.}{P{\^a}ris
  et~al.}{2017}]{Paris2016}
P{\^a}ris I.,  et~al., 2017, \mn@doi [A\&A] {10.1051/0004-6361/201527999}, 597,
  A79

\bibitem[\protect\citeauthoryear{{P\'erez-R\`afols} et~al.,}{{P\'erez-R\`afols}
  et~al.}{2018}]{Perez2018}
{P\'erez-R\`afols} I.,  et~al., 2018, \mn@doi [MNRAS] {10.1093/mnras/stx2525},
  \href {http://adsabs.harvard.edu/abs/2018MNRAS.473.3019P} {473, 3019}

\bibitem[\protect\citeauthoryear{{Pieri}}{{Pieri}}{2014}]{Pieri2014}
{Pieri} M.~M.,  2014, \mn@doi [MNRAS] {10.1093/mnrasl/slu142}, \href
  {http://adsabs.harvard.edu/abs/2014MNRAS.445L.104P} {445, L104}

\bibitem[\protect\citeauthoryear{{Rauch}}{{Rauch}}{1998}]{Rauch1998}
{Rauch} M.,  1998, \mn@doi [ARA\&A] {10.1146/annurev.astro.36.1.267}, \href
  {http://adsabs.harvard.edu/abs/1998ARA%26A..36..267R} {36, 267}

\bibitem[\protect\citeauthoryear{{Rauch}, {Haehnelt}  \& {Steinmetz}}{{Rauch}
  et~al.}{1997}]{Rauch1997}
{Rauch} M.,  {Haehnelt} M.~G.,   {Steinmetz} M.,  1997, \mn@doi [ApJ]
  {10.1086/304085}, \href {http://adsabs.harvard.edu/abs/1997ApJ...481..601R}
  {481, 601}

\bibitem[\protect\citeauthoryear{{Richards} et~al.,}{{Richards}
  et~al.}{2009}]{Richards2009}
{Richards} G.~T.,  et~al., 2009, \mn@doi [ApJS] {10.1088/0067-0049/180/1/67},
  \href {http://adsabs.harvard.edu/abs/2009ApJS..180...67R} {180, 67}

\bibitem[\protect\citeauthoryear{Ross et~al.}{Ross et~al.}{2009}]{Ross2009}
Ross N.~P.,  et~al., 2009, \mn@doi [ApJ] {10.1088/0004-637X/697/2/1634}, 697,
  1634

\bibitem[\protect\citeauthoryear{{Ross} et~al.,}{{Ross}
  et~al.}{2012}]{Ross2012}
{Ross} N.~P.,  et~al., 2012, \mn@doi [ApJS] {10.1088/0067-0049/199/1/3}, \href
  {http://adsabs.harvard.edu/abs/2012ApJS..199....3R} {199, 3}

\bibitem[\protect\citeauthoryear{{Sargent}, {Young}, {Boksenberg}  \&
  {Tytler}}{{Sargent} et~al.}{1980}]{Sargent1980}
{Sargent} W.~L.~W.,  {Young} P.~J.,  {Boksenberg} A.,   {Tytler} D.,  1980,
  \mn@doi [ApJS] {10.1086/190644}, \href
  {http://adsabs.harvard.edu/abs/1980ApJS...42...41S} {42, 41}

\bibitem[\protect\citeauthoryear{{Schaye}, {Aguirre}, {Kim}, {Theuns}, {Rauch}
  \& {Sargent}}{{Schaye} et~al.}{2003}]{Schaye2003}
{Schaye} J.,  {Aguirre} A.,  {Kim} T.-S.,  {Theuns} T.,  {Rauch} M.,
  {Sargent} W.~L.~W.,  2003, \mn@doi [ApJ] {10.1086/378044}, \href
  {http://adsabs.harvard.edu/abs/2003ApJ...596..768S} {596, 768}

\bibitem[\protect\citeauthoryear{Shen et~al.}{Shen et~al.}{2007}]{Shen2006}
Shen Y.,  et~al., 2007, \mn@doi [Astron. J.] {10.1086/513517}, 133, 2222

\bibitem[\protect\citeauthoryear{Shen et~al.}{Shen et~al.}{2009}]{Shen2008}
Shen Y.,  et~al., 2009, \mn@doi [ApJ] {10.1088/0004-637X/697/2/1656}, 697, 1656

\bibitem[\protect\citeauthoryear{{Slosar} et~al.,}{{Slosar}
  et~al.}{2011}]{Slosar2011}
{Slosar} A.,  et~al., 2011, \mn@doi [JCAP] {10.1088/1475-7516/2011/09/001},
  \href {http://adsabs.harvard.edu/abs/2011JCAP...09..001S} {9, 001}

\bibitem[\protect\citeauthoryear{{Slosar} et~al.,}{{Slosar}
  et~al.}{2013}]{Slosar2013}
{Slosar} A.,  et~al., 2013, \mn@doi [JCAP] {10.1088/1475-7516/2013/04/026},
  \href {http://adsabs.harvard.edu/abs/2013JCAP...04..026S} {4, 026}

\bibitem[\protect\citeauthoryear{Tinker, Robertson, Kravtsov, Klypin, Warren,
  Yepes  \& Gottlober}{Tinker et~al.}{2010}]{Tinker2010}
Tinker J.~L.,  Robertson B.~E.,  Kravtsov A.~V.,  Klypin A.,  Warren M.~S.,
  Yepes G.,   Gottlober S.,  2010, \mn@doi [ApJ] {10.1088/0004-637X/724/2/878},
  724, 878

\bibitem[\protect\citeauthoryear{{Vikas} et~al.,}{{Vikas}
  et~al.}{2013}]{Vikas2013}
{Vikas} S.,  et~al., 2013, \mn@doi [ApJ] {10.1088/0004-637X/768/1/38}, \href
  {http://adsabs.harvard.edu/abs/2013ApJ...768...38V} {768, 38}

\bibitem[\protect\citeauthoryear{{Weymann}, {Morris}, {Foltz}  \&
  {Hewett}}{{Weymann} et~al.}{1991}]{weymann1991}
{Weymann} R.~J.,  {Morris} S.~L.,  {Foltz} C.~B.,   {Hewett} P.~C.,  1991,
  \mn@doi [ApJ] {10.1086/170020}, \href
  {http://adsabs.harvard.edu/abs/1991ApJ...373...23W} {373, 23}

\bibitem[\protect\citeauthoryear{White et~al.}{White et~al.}{2012}]{White2012}
White M.,  et~al., 2012, \mn@doi [MNRAS] {10.1111/j.1365-2966.2012.21251.x},
  424, 933

\bibitem[\protect\citeauthoryear{{Y{\`e}che} et~al.,}{{Y{\`e}che}
  et~al.}{2010}]{Yeche2010}
{Y{\`e}che} C.,  et~al., 2010, \mn@doi [A\&A] {10.1051/0004-6361/200913508},
  \href {http://adsabs.harvard.edu/abs/2010A%26A...523A..14Y} {523, A14}

\bibitem[\protect\citeauthoryear{{York} et~al.,}{{York}
  et~al.}{2000}]{York2000}
{York} D.~G.,  et~al., 2000, \mn@doi [AJ] {10.1086/301513}, \href
  {http://adsabs.harvard.edu/abs/2000AJ....120.1579Y} {120, 1579}

\bibitem[\protect\citeauthoryear{{du Mas des Bourboux} et~al.,}{{du Mas des
  Bourboux} et~al.}{2017}]{Helion2017}
{du Mas des Bourboux} H.,  et~al., 2017, \mn@doi [A\&A]
  {10.1051/0004-6361/201731731}, \href
  {http://adsabs.harvard.edu/abs/2017A%26A...608A.130D} {608, A130}

\makeatother
\end{thebibliography}
\bibliographystyle{mnras}

\end{document}